\documentclass[aps,pre,twocolumn,groupedaddress,showpacs]{revtex4-1}

\usepackage{amssymb,amsfonts,amsmath}
\usepackage{bm,graphicx}
\usepackage{color}
\usepackage[dvips]{epsfig}
\usepackage{longtable}

\newcommand{\boldsig}{\mbox{\boldmath{$\sigma$}}}
\newcommand{\boldeps}{\mbox{\boldmath{$\epsilon$}}}

\begin{document}

\title[]{Spatiotemporal complexity of electroconvection patterns in nematic liquid crystals} 

%Authors
\author{Alexei Krekhov}
\email[]{alexei.krekhov@ds.mpg.de}
\affiliation{Max Planck Institute for Dynamics and Self-Organization, 37077 G\"ottingen, Germany}

\author{Bernd Dressel}
\author{Werner Pesch}
\affiliation{Physikalisches Institut, Universit\"at Bayreuth, 95440 Bayreuth, Germany}

\author{Vladimir Delev}
\author{Eduard Batyrshin}
\affiliation{Institute of Molecule and Crystal Physics, Ufa Research Center, Russian Academy of Sciences, 450075 Ufa, Russia}

\date{\today}

%%% Abstract
\begin{abstract}
We investigate a number of complex patterns driven by the electro-convection instability in a planarly aligned layer of a nematic liquid crystal.
They are traced back to various secondary instabilities of the ideal roll patterns bifurcating at onset of convection, whereby the basic nemato-hydrodynamic equations are solved by common Galerkin expansion methods.
Alternatively these equations are systematically approximated by a set of coupled amplitude equations.
They describe slow modulations of the convection roll amplitudes, which are coupled to a flow field component with finite vorticity perpendicular to the layer and to a quasi-homogeneous in-plane rotation of the director.
It is demonstrated that the Galerkin stability diagram of the convection rolls is well reproduced by the corresponding one based on  the amplitude equations.
The main purpose of the paper is, however, to demonstrate that their direct numerical simulations match surprisingly well new experiments, which serves as a convincing test of our theoretical approach.
\end{abstract}

%\keywords{}

\maketitle

% Body of paper here - Use proper section commands

%
%%%% Introduction %%%%
%
\section{\label{intro} Introduction}
Electroconvection (EC) in the planar configuration of nematic liquid crystals is a prime paradigm for pattern forming instabilities in anisotropic systems (see, e.g., \cite{Bodenschatz:1988,Buka:1996,Golovin:2006}).
Nematic liquid crystals (nematics) are fluids with a long range orientational ordering of their nonspherical molecules, the mean orientation of which is described by the director field $\bm n = (n_x, n_y, n_z)$ subject to the normalization ${\bf n}^2=1$.
In the standard, {\it planar}, experimental setup a nematic layer of thickness $d$, $10$~$\mu$m~$\lesssim d \lesssim 50$~$\mu$m, is sandwiched between two electrically conducting plates parallel to the $x,y$ plane.
By an appropriate treatment of their surfaces the director is homogeneously oriented parallel to the plates along a fixed direction (along $\bm{\hat x}$).
A sinusoidal ac voltage of angular frequency $\omega$ with the root mean square (rms) amplitude $U_0(\omega)$ is then applied transversely (along $\bm{\hat z}$).
If $U_0(\omega)$ exceeds a certain critical value $U_c(\omega)$ the homogeneous basic state is destabilized against the EC instability.
One observes a common convection roll pattern whose periodicity in the $x,y$ plane is characterized by the critical wave vector $\bm q_c(\omega)$.
Due to the associated periodic out-of-plane modulation of the director ($n_z \ne 0$) in EC the nematic layer acts as an optical grating, such that the patterns are easily visualized.
Though there exists some analogy of EC to isotropic thermal Rayleigh-B\'enard convection (RBC), the present system is generically anisotropic in the plane of the nematic layer.
As a consequence, the angle between $\bm q_c$ and the preferred $x$ axis is small and in most cases even zero for moderate $\omega$.
The richness of the various EC scenarios derives from the fact that the ac frequency $\omega$ serves as an important second control parameter.
Apart from the $\omega$ dependence of $\bm q_c$ and $U_c$, also certain time-symmetries of the convection patterns and their secondary instabilities with increasing voltage depend on $\omega$.
A big advantage is that the characteristic times in EC are typically short, which allows one to record the pattern dynamics during relatively short measurements.
Furthermore, one can easily achieve large aspect ratios in experiments, i.e., the horizontal extension of the patterns is much larger than the roll diameter.
In fact one may observe quasi-uniform roll patterns of up to $500$ rolls.
Consequently, it is safe to use in the theoretical analysis periodic boundary conditions in the plane of the nematic layer.
That leads to essential simplifications by switching from position to Fourier space with respect to the horizontal spatial coordinates.
A full theoretical description of EC starting from the well accepted nematohydrodynamic equations (see, e.g., \cite{Gennes:1974,chandra:1977,brand:1996}), which describe the intricate coupling of the electric field $\bm E$, of the director $\bm n$, and of the flow field $\bm v$, is highly demanding.
Compared to isotropic RBC one has to deal with quintic nonlinearities and with the time dependence of the applied voltage, apart from the additional director field.
Furthermore, the primary roll patterns show already quite near to the EC onset secondary and tertiary bifurcations to complex spatio-temporal planforms, characterized for instance by a persistent generation and annihilation of defects like dislocations or grain boundaries.
The various nematics differ in their material parameters (e.g., electric conductivities, dielectric constants), which may also have a considerable impact on the pattern morphology.
In this paper we concentrate exclusively on the material MBBA (N-4-methoxybenzylidene-4-butylaniline).
Already in the past it has been used in many experimental and theoretical EC studies, in particular since the MBBA material parameters are fairly well known.
Our theoretical analysis of EC starts as usual with the determination  of the critical properties, $U_c(\omega)$, $\bm q_c(\omega)$, at onset of convection.
Then we construct an exact ``Galerkin'' stability diagram of the  resulting roll patterns in the $\omega - U_0$ plane.
The calculations make use of specific series expansions, which are adapted to the assumed periodic boundary conditions of $\bm E$, $\bm n$, $\bm v$ in the $x,y$ plane and to the vertical boundary conditions at $z =\pm d/2$, where for instance $\bm v$ has to vanish.
The main theme of the present work is the description of the complex dynamics of the pattern that develops at the stability boundaries of the  stability diagram.
The crucial point is that the theoretical analysis of the nemato-hydrodynamic equations (NHE) can be substantially simplified by reducing them to a system of partial differential (amplitude) equations in the horizontal coordinates; a general discussion of this procedure is for instance found in \cite{Cross:1993,CroGreen:2009}.
The first amplitude equations for EC have been presented in \cite{Bodenschatz:1988}, which reflect clearly the anisotropy of the system.
Even the complex dynamics of topological defects (dislocations) in the roll pattern has been assessed \cite{Bodenschatz2:1988,Bodencom:1990}.
Later investigations have concentrated on the proper description of the important secondary instabilities in EC in the form of long-wavelength modulations of the roll orientations, which have been first described in RBC (see, e.g., \cite{Busse:1979}).
Such instabilities are reinforced by an induced flow field component with vertical vorticity associated with any roll curvature, which has thus been incorporated into the amplitude equations in \cite{Kaiser:1993} (for the analogous approach in RBC, see \cite{Decker:1994}).
Even these generalized amplitude equations were still incomplete as demonstrated clearly in \cite{Plaut:1997}.
The key result was that a {\it twist} distortion of $\bm n$ in the form of a {\it homogeneous rotation} (i.e., independent of $x,y$) in the plane opens a novel route for a destabilization of rolls in the weakly nonlinear regime.
This leads to the so-called {\it abnormal} convection rolls, whose director 
field contains not only the usual, in $x,y$ periodic, out-of-plane distortion 
$n_z \ne 0$, but also the finite twist component $n_y$.
For completeness, it should be mentioned, that the study of thermally driven roll patterns in planar nematics has revealed clear analogies to EC patterns (see, e.g., \cite{Plaut:1999}).
Thus it is not surprising that homogeneous director modes are also important in this system \cite{Dressel:2002}.
To describe the abnormal rolls and their instabilities, it was necessary to generalize the amplitude equations further by including the twist mode.
This task was accomplished in \cite{Plaut:1999}, where a first version of the resulting system of three coupled amplitude equations (CAE) was presented.
In the present paper we have derived a more general version of the CAE, by which many features of our exact Galerkin stability diagram analysis are well reproduced; this serves as a first important test of the reliability of the CAE approximation of the full NHE.
To validate the CAE concept in further detail, the EC instability has been systematically studied in new experiments with MBBA.
When varying frequency and amplitude of the applied ac voltage, one is confronted with a number of different complex experimental convection patterns, which match surprisingly well the corresponding numerical simulations of the CAE.
The paper is organized as follows.
In Sec.~\ref{sec:basic_eq} we discuss briefly the underlying nematohydrodynamic equations and the methods to calculate from them the convection roll patterns and their stability.
In Sec.~\ref{sec:experim} the experimental setup, which was used for our new experimental studies, is briefly sketched.
The Galerkin stability diagram of rolls near onset is discussed in Sec.~\ref{sec:stabdiag}.
Section~\ref{sec:Ampl_Eqs} is devoted to the derivation and the discussion of the CAE.
Selected simulations of the CAE together with a detailed comparison with the experiments are presented in Sec.~\ref{sec:simul}.
The paper concludes with some final remarks in Sec.~\ref{sec:summary}.
In the Appendixes we list the material parameters of MBBA together with the coefficients of the amplitude equations as used in this paper.
%

%
%%%%% Basic equations %%%%%
%
\section{Theoretical analysis of EC}
\label{sec:basic_eq}
The nematic fluids used in EC experiments contain a small amount of mobile ions (either associated with the production process or brought in by doping) which leads to a very small electric conductivity of the scale $\overline{\sigma}_{0}$ typically between $10^{-9}$~$(\Omega$~m)$^{-1}$ and $10^{-7}$~$(\Omega$~m)$^{-1}$.
Due to uniaxial symmetry of nematics all constitutive equations are of tensorial nature.
Thus the electric current density $\bm j$ and the dielectric displacement $\bm D$ are related to the electric field $\bm E$ as follows:
\begin{eqnarray}
\label{eq:P}
\bm{j} = \overline{\sigma}_{0} \boldsig \bm{E} \,, \;
\bm{D} = \epsilon_0 \boldeps \bm{E} \,,
\end{eqnarray}
with the dimensionless tensors \boldsig\ (electrical conductivity), \boldeps\ (dielectric permittivity) and the vacuum permittivity $\epsilon_0 = 8.8542 \times 10^{-12}$~A~s/(V~m).
The electric field is represented as:
\begin{equation}
\label{eq:phi}
\bm E(x,y,z,t) = E_0 \cos(\omega t) \bm{\hat z} - \nabla \phi(x,y,z,t) \,, \;
E_0 =\sqrt{2} U_0 / d \,,  
\end{equation}
where the electric potential $\phi$ describes the correction of the applied electric ac field $\propto E_0$ inside the nematic layer in the presence of convection.
As standard for any substance with uniaxial symmetry the components of the tensors \boldsig\ and \boldeps\ are represented in terms of the director components $n_i$ with $i= x,y,z$ as follows:
\begin{eqnarray}
\label{eq:eps_sigma}
\sigma_{ij} = \sigma_\perp \delta_{ij} + \sigma_a n_i n_j \,, \;
\epsilon_{ij} = \epsilon_\perp \delta_{ij} + \epsilon_a n_i n_j \,.
\end{eqnarray}
The strength of the anisotropies is quantified by the dimensionless parameters $\sigma_a = \sigma_{\parallel} - \sigma_{\perp}$ and $\epsilon_a = \epsilon_{\parallel} - \epsilon_{\perp}$.
For instance when $\bm E, \bm j \perp \bm n$ the conductivity is given as $\overline{\sigma}_{0} \sigma_{\perp}$ while for $\bm E, \bm j \parallel \bm n$, it is given as $\overline{\sigma}_{0} \sigma_{\parallel}$.
The signs of $\sigma_a$, $\epsilon_a$, which can be both positive and negative for the various nematic materials, play a key role to understand the possible driving mechanism for EC (see, e.g., \cite{Golovin:2006}); for MBBA one finds $\sigma_a >0$, $\epsilon_a <0$.
Since flexoelectric effects (see, e.g., \cite{Krekhov:2011}) turn out to be unimportant we use a simplified version of the general NHE, the so-called standard model \cite{Bodenschatz:1988}, where the potential $\phi$ is determined by the Maxwell equations in the quasi-static approximation (charge conservation).
The velocity field $\bm v$ is determined by a (generalized) Navier-Stokes equation in the presence of the volume force $\rho_{el} \bm E$ with the charge density $\rho_{el} = \nabla \cdot \bm D$.
The anisotropic stress tensor depends in a complicated manner on $\bm v$, $\bm n$ and their gradients.
Five independent viscosity coefficients $\alpha_i$, $i =1 \dots 5$ come into play, which are of the order of $\alpha_0 = 10^{-3}$~Pa~s.
The dynamics of the director $\bm n$ is governed by the balance of electric, viscous and orientational elastic torques on the director.
The latter depend on three additional elastic constants $k_{ii}$, $i =1 \dots 3$ of the order of $k_0 =10^{-12}$~N.
%

%%%
\subsection{Non-dimensionalization of the basic equations}
\label{subsec:nondim}
The number of parameters appearing in the standard model can be reduced to some extent by introducing suitable dimensionless quantities.
Lengths are measured in units of $d/\pi$, time in units of the director relaxation time $\tau_d = \alpha_0 d^2/(k_0 \pi^2)$, the elastic constants in units of $k_0$ and the viscosity coefficients in units of $\alpha_0$.
It turns out that the dimensionless ratio $Q = \tau_d/{\tau_q}$ plays an important role, where $\tau_q =\epsilon_0 \epsilon_{\perp} / (\overline{\sigma}_{0} \sigma_\perp)$ denotes the charge relaxation time.
For large $Q$ the conductivity scale $\overline{\sigma}_0$ appears only in the dimensionless frequency parameter $\omega^\prime = \omega \tau_q$, which also simplifies considerably the interpretation of the experimental data.
Large $Q$ require obviously fairly large values of $d$ together with not too small $\overline{\sigma}_0$.
Since a direct measurement of $\overline{\sigma}_0$ in convection cells is not trivial, $\overline{\sigma}_0$ is typically used as fit parameter to match the frequency scale in theory and experiments.
For convenience we have listed all material parameters of the nematic MBBA in Appendix~\ref{sec:appMBBA}, where one finds also the values of $\tau_d$, $\tau_q$ to be used in this paper. 
The main control parameter $U_0$ is typically parametrized by the dimensionless quantity $R$ or by the relative distance $\varepsilon$ to the onset of convection, which are defined as follows:
\begin{eqnarray}
\label{eq:R}
R = \frac{\epsilon_0 2 U^2_0}{k_0 \pi^2} \,, \;
\varepsilon = \frac{R-R_c}{R_c} 
\quad \textrm{with} \;\;
R_c  = \frac{\epsilon_0 2 U^2_c}{k_0 \pi^2} \,.
\end{eqnarray}
%

%%%
\subsection{EC roll solutions and their stability}
\label{subsec:stab}
In general the roll solutions of the NHE are determined using Galerkin methods.
The spatial periodicity of all fields with respect to the $x,y$ coordinates, which is governed by a wave vector $\bm q =(q,p)$, is captured by 2D Fourier series.
With respect to the $z$ dependence we expand into complete sets of functions which fulfill the boundary conditions at the confining plates ($z = \pm d/2$).
Here the director is parallel to the $x$ direction, while the induced electrical potential $\phi$ as well as the velocity field $\bm v$ have to vanish there.
For instance, we use for the $z$ component of $\bm n$ the following expansion:
\begin{equation}
\label{eq:nz}
n_z(\bm x, z, t) = \sum^K_{k = -K} \sum^M_{m = 1}
\overline{n}_z(\bm q,t; k, m) \exp ( i \,k \, \bm q \cdot \bm x ) 
S_m(z) \,,
\end{equation}
with $S_m(z) = \sin[m \pi (z/d + 1/2)]$ and $\bm x = (x,y)$.
Analogous truncated series expansions are used for $\phi$, for $\bm v$ and for $n_x$, $n_y$.
In this way the NHE are mapped to a system of coupled ordinary differential equations (ODEs) in time $t$ for the respective expansion coefficients.
The standard analysis of these equations in the linear regime shows that steady nontrivial spatially periodic roll solutions with wave vector $\bm q$ exist only when the control parameter $R$ [Eq.~(\ref{eq:R})] fulfills the condition $R > R_0(\bm q)$ with so-called neutral surface $R_0(\bm q)$.
The minimum of $R_0(\bm q)$ with respect to $\bm q$ defines the critical wave vector $\bm q_c$ and the critical control parameter $R_c = R_0(\bm q_c)$, meaning that the basic homogenous planar state undergoes at $R = R_c$ the EC convection instability.
It turns out that both $R_c$ and $|\bm q_c|$ are monotonically increasing functions of $\omega$.
Returning from $R_c$, $\bm q_c$ to physical units the critical rms voltage $U_0 =U_c$ would vary between $10$~V and $100$~V and $|\bm q_c| \sim O(1/d)$, respectively.
The bifurcation to EC at $R = R_c$ is supercritical, i.e., steady roll solutions exist only for $R \gtrsim R_c$, i.e., for positive $\varepsilon =(R - R_c)/R_c$ [Eq.~(\ref{eq:R})].
To determine the roll solutions in the case of purely sinusoidal ac driving voltage as assumed in this paper it is convenient to expand in addition all expansion coefficients like $\overline{n}_z(\bm q, t; k, m)$ in Eq.~(\ref{eq:nz}) as truncated Fourier series in time as follows:
\begin{eqnarray}
\label{eq:nzfour}
\overline{n}_z (\bm q, t; k, m ) =
\sum^N_{n = -N} 
\hat n_z(\bm q; n, k, m) \exp ( i \,n \, \omega t ) \,.
\end{eqnarray}
In principle we should also envisage a subharmonic response of our parametrically driven system by summing in Eq.~(\ref{eq:nzfour}) over half integers $n/2$ as well.
This possibility is in particular realized when the applied voltage is characterized by two different frequencies \cite{John:2004,Pietschmann:2010}.
However, for pure sinusoidal ac voltage the ansatz Eq.~(\ref{eq:nzfour}) turns out to be sufficient as the consequence of special symmetries of the NHE with respect to a time shift by half period $\pi/\omega$ in combination with the reflection $z \rightarrow -z$.
At the end, the NHE are mapped to a system of nonlinear algebraic equations for the expansion coefficients $\hat n_z(\bm q; n, k, m)$ in Eq.~(\ref{eq:nzfour}) and the corresponding ones for the other fields, which is solved by the Newton-Raphson method.
The tedious mapping procedure has been fully automatized using {\it Mathematica} to produce finally a number of Fortran codes, which are exploited in this work.
Compared to previous calculations (see, e.g., \cite{Plaut:1999}) the numerical effort has been substantially reduced.
In particular the annoying quintic nonlinearities of the original NHE could be replaced by cubic ones by introducing certain products of director components as new auxiliary variables.
We are interested in the stability diagram of rolls with $\bm q = \bm q_c$, i.e., in the regions in the $\omega - \varepsilon$ plane where the rolls are stable.
For that purpose the coupled ordinary differential equations for the time-dependent expansion coefficients like $\overline{n}_z(\bm q_c, t; k, m)$ in Eq.~(\ref{eq:nz}) are linearized about the respective roll solutions in terms of their linear perturbations.
Thus one arrives at a set of linear ODEs with time-periodic coefficients which are solved with the use of a standard Floquet ansatz.
For instance, the perturbation $\delta \overline{n}_z(\bm q_c, t; k, m)$ of $\overline{n}_z(\bm q_c, t; k, m)$ is thus represented as:
\begin{eqnarray}
\label{eq:delt_nz}
\delta \overline{n}_z(\bm q_c, t; k, m) &&=
\exp(\sigma t) \exp( i \, \bm s \cdot \bm x ) 
\nonumber \\
&&\times \sum^N_{n = -N} 
\delta \hat n_z(\bm q_c; n, k, m) \exp( i \,n \, \omega t ) \,. \;\;
\end{eqnarray}
with the Floquet vector $\bm s$.
In this way we arrive for fixed  $\bm q_c(\omega)$ at a linear eigenvalue problem with a set of eigenvalues $\sigma_j(\varepsilon, \bm s)$, $j=1,2,\dots$ where $Re(\sigma_1) \gtrsim Re(\sigma_2), \dots$.
Next we define a function $\varepsilon^0(\bm s)$, such that for each $\bm s$ the smallest $\varepsilon$, which solves $Re[\sigma_1 (\varepsilon, \bm s)] =0$ is given as $\varepsilon = \varepsilon^0(\bm s)$.
Consequently the primary rolls become unstable at the minimum $\bm s = \bm s_0$ of $\varepsilon^0(\bm s)$, i.e., at $\varepsilon_{inst} = \varepsilon^0(\bm s_0)$.
Important features of the EC stability diagram are governed by long-wavelength instabilities with $|\bm s_0| \ll q_c$.
They are expressed as slow spatial modulations of the rolls patterns, which have been thoroughly studied before in RBC (see, e.g., \cite{Busse:1979}).
The special case $\bm s_0 \perp \bm{q}_c$ is known as the zig-zag (ZZ) instability, whereas the skewed varicose (SV) instability stands for the more general case of an arbitrary angle between $\bm s_0$ and $\bm q_c$.
These instabilities are reflected in the average $\langle n_z(\bm x) \rangle$ of $n_z(\bm x,z,t)$ with respect to $z$ and $t$ which is for $\varepsilon \gtrsim \varepsilon_{inst}$ well approximated by:
\begin{eqnarray}
\label{eq:longwav} 
\langle n_z(\bm x) \rangle &&= 
A \cos(\bm q \cdot \bm x) [1  + B \sin (\bm s_0 \cdot \bm x)]
\quad \textrm{with}
\nonumber \\
&&A \propto \sqrt{\varepsilon} \,, \; 
B \propto \sqrt{ \varepsilon - \varepsilon_{inst}} \,.
\end{eqnarray}
The in-plane rotation or ``twist'' of the director gives rise to the additional destabilization mechanisms.
The corresponding perturbation $\delta n_y$ has a finite time average and is even in $z$.
%

%
%%%% Experimental setup %%%%
%
\section{Experimental setup}
\label{sec:experim}
In this paper we refer mainly to new systematic experiments with the nematic liquid crystal MBBA (TCI Co.).
To work in a convenient range of ac frequencies up to $f \sim 1000$~Hz the material was weakly doped with tetra-ethyl-ammonium chloride.
The nematic is then filled into a commercially available cell (EHC Co., Japan) with ITO electrodes, which are coated with rubbed polymer coatings to provide the planar alignment of the nematic.
The cell thickness was $d = (25 \pm 1)$~$\mu$m and the temperature of the cell was kept at $(26 \pm 0.1)$~$^\circ$C, which is well adapted to the material parameter set listed in Appendix~\ref{sec:appMBBA}.
The sinusoidal voltage $U(t) =\sqrt{2} U_0 \sin(\omega t)$ applied to the cell was provided by an Agilent~33220 waveform generator and a Tabor~9200 voltage amplifier.
To determine the onset of convection the value of $U_0$ is slowly increased at fixed $\omega$.
Typically we changed the voltage every $3 - 5$~min by steps between $0.01 - 0.1$~V.
To map out the whole stability diagram and to study the patterns above EC threshold the waiting time between successive voltage steps has been increased to $10 - 30$~min, such that transient dynamical processes  would have died out.
The EC patterns act as an optical grating, which allows their detailed exploration by shining light through the convection cell (for a most recent theoretical analysis, see \cite{Trainoff:2002,Pesch:2013} and references therein).
The light coming from an illuminator passes a polarizer before entering the cell and an analyzer above the cell.
We used two different optical setups (O1, O2) to enhance the impact of specific director components to the pattern images.
(O1) Both the polarizer and the analyzer are oriented parallel to the initial planar director orientation ($\parallel \bm{\hat x}$).
This geometry is in particular sensitive to the out-of-plane distortion, $n_z$, of the director field, which is exploited in the standard  shadowgraph analysis of EC patterns.
In this way the basic periodicity of the roll patterns bifurcating at onset of convection, i.e., at $R =R_c$ with wave vector $\bm q_c = (q_c,0)$, is immediately visible.
For $R > R_c$ one easily observes amplitude and phase modulations in these patterns, which can be traced back to the secondary long-wavelength instabilities of rolls described in Sec.~\ref{sec:basic_eq}. 
(O2) The polarizer is oriented perpendicular to the initial planar director orientation, while the analyzer remains parallel to it; furthermore an additional quarter-wave plate is placed between the cell and the analyzer.
This geometry is in particular sensitive to the identification of the $n_x$, $n_y$ components of the director, i.e., to the in-plane twist of the director \cite{Rudroff:1999,Amm:1999,Oikawa:2004}, and thus to the spatial variations of the twist amplitude $\varphi \propto \arctan (n_y/n_x)$.
The EC patterns are visualized with a Zeiss Axio Imager A1m polarizing microscope together with a Zeiss HAL~100 halogen white light illuminator.
The resulting images were recorded by an Optronis CL600x2 camera with a spatial resolution of 512$\times$512 pixels and 256 gray levels.
All images were normalized to a  background image taken slightly below onset of electroconvection to minimize the effects of inhomogeneities in the optical system and in the experimental cell.
%

%%%% EC stability diagram %%%%
%
\section{Electroconvection stability diagram for MBBA}
\label{sec:stabdiag}
\begin{figure}[ht]
\centering
\includegraphics[width=0.95\linewidth]{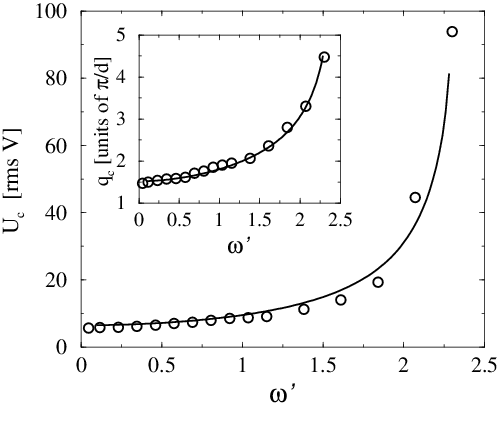}
\caption{\label{CRITMBBA}
Critical rms voltage $U_c$ and critical wave number $q_c$ for MBBA as function of $\omega^\prime = \omega \tau_q$ in the conductive regime from theory (solid lines) in comparison with experimental data (circles).}
\end{figure}

In the following we discuss the stability diagram of EC rolls for MBBA (see Appendix~\ref{sec:appMBBA}) near onset of convection in the $\omega - R$ plane.
The calculation makes use of the theoretical tools introduced in Sec.~\ref{sec:basic_eq} according to which the number of the algebraic equations to be solved is determined by the truncation parameters $M$, $N$, $K$ for instance in the ansatz for the director component $n_z$ [Eq.~(\ref{eq:nz})].
Their values used for the different calculations in this paper will be explicitly given, where we made sure that increasing these values does not lead to visible modifications in the graphical presentations of our theoretical results.
It will be demonstrated, that our system is in fact characterized by a large value of $Q =\tau_d /\tau_q$, such that $\omega$ appears only in the form $\omega^\prime = \omega \tau_q$ (see Sec.~\ref{subsec:nondim}).
To determine the onset of convection ($U_c, \bm q_c$) the cutoffs $K =1$, $M =4$, $N =1$ have been used.
We find for all $\omega$ only a bifurcation to {\it normal rolls} at $R = R_c$ with $\bm q_c \parallel \bm{\hat x}$ where $n_y \equiv 0$.
A closer look reveals, however, two roll types.
For $\omega \lesssim \omega_c$, with the cutoff-frequency $\omega_c$, in the so-called conductive regime, the time average of $n_z(\bm x, z, t)$ [Eq.~(\ref{eq:nz})] vanishes in leading order.
This means that $\hat n_z(\bm q_c; n=0, k=m=1)$ in Eq.~(\ref{eq:nzfour}) is finite, whereas $\hat n_z(\bm q_c; n=1, k=m=1) =0$.
For $\omega \gtrsim \omega_c$ in the so-called dielectric regime $n_z(\bm x, z, t)$ oscillates with the ac frequency, which implies finite expansion coefficients $\hat n_z(\bm q_c; n=1, k=m=1)$, while $\hat n_z(\bm q_c; n=0, k=m=1) =0$.
In this paper we will restrict ourselves to the conductive regime.
The resulting curves for $U_c$ and $q_c$ as function of the dimensionless frequency $\omega^\prime = \omega \tau_q$ are shown in Fig.~\ref{CRITMBBA}, where $\omega^\prime_c = 2.3$.
In the experiments we find $\omega_c = 2 \pi f_c$ with $f_c \approx 980$~Hz, which yields $\omega^\prime_c = 2.3$ as well by using $\tau_q \approx 0.37 \times 10^{-3}$~s [corresponding to  $\overline{\sigma}_0 =1.25 \times 10^{-7}$~$(\Omega$~m)$^{-1} \equiv \overline{\sigma}_{exp}$], as fit parameter.
Inspection of all the experimental data for $U_c$, $q_c$ in Fig.~\ref{CRITMBBA} shows their excellent agreement with the theory in the whole $\omega^\prime$ range.
In the following we will hold on $\overline{\sigma}_{exp}$ to characterize in 
general the conductivity scale of MBBA in the experiments, which leads indeed 
to large $Q \approx 170$ (see Appendix~\ref{sec:appMBBA}).
For completeness it should be mentioned that our analysis does not cover the case of very small $\omega$ of the order of $10^{-3}$~s$^{-1}$, where $\omega \tau_d$ becomes small.
In this case the director dynamics looks spiky \cite{Krekhov:2011,Eber:2012} and one would need many Fourier modes in time ($N \gg 1$) to resolve it.
Thus we exclude in this paper very low $\omega$ at all.
%

%%% Figure 1
%

%%% Figure 2
\begin{figure}[ht]
\centering
\includegraphics[width=0.95\linewidth]{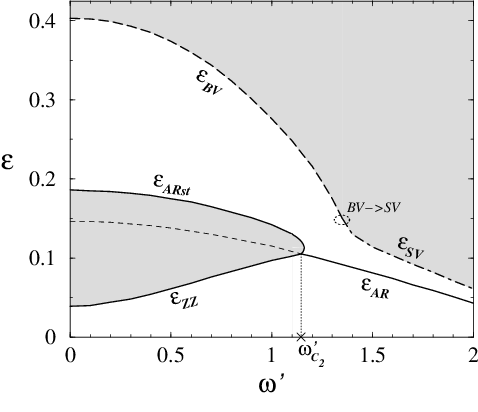}
\caption{\label{fig:STABMBBA}
Stability diagram for normal rolls in MBBA in the $\omega^\prime - \varepsilon$ plane for the same parameters as in Fig.~\ref{CRITMBBA}.}
\end{figure}
The stability diagram of normal EC rolls with wave vector $\bm q_c =(q_c,0) \parallel \bm{\hat x}$ in the $\varepsilon - \omega^\prime$ plane is shown in Fig.~\ref{fig:STABMBBA} for the same material  parameters as in Fig.~\ref{CRITMBBA} and for the cutoff parameters  $K =3$, $M =4$, and $N =2$.
First, the normal roll solutions in the nonlinear regime with $\varepsilon > 0$ are calculated following the procedure sketched in Sec.~\ref{subsec:stab}.
Above the line $\varepsilon =\varepsilon_{AR}$ they become unstable against a homogeneous rotation of the director in the plane.
Consequently, the stability boundaries in Fig.~\ref{fig:STABMBBA} for $\varepsilon >\varepsilon_{AR}$ refer always to the resulting abnormal roll solutions with a finite $n_y$ component.
In the lower gray-shaded regions for $\omega^\prime < \omega^\prime_{C_2} \approx 1.15$, where $\omega^\prime_{C_2}$ defines a kind of codimension-2 point, the normal rolls are destabilized already for $\varepsilon < \varepsilon_{AR}$ against the long-wavelength zig-zag (ZZ) instability discussed after Eq.~(\ref{eq:delt_nz}).
For $\varepsilon > \varepsilon_{AR}$ the basic abnormal rolls remain unstable against a skewed-varicose (SV) instability until the restabilization line $\varepsilon = \varepsilon_{ARst}$ is crossed.
The angle between the Floquet vector $\bm s_0$ and $\bm q_c$ is here about $70^\circ$ and thus slightly smaller then $90^\circ$ for the zig-zag instability.
Increasing $\varepsilon$ further the abnormal EC rolls remain stable until the so-called bimodal instability line $\varepsilon = \varepsilon_{BV}$.
Here the rolls become unstable against convection rolls with a wave vector $\bm q$ with $|\bm q| = O(q_c)$ including a finite angle with $\bm q_c$.
This instability will not be discussed in this paper; some details can be found in \cite{Plaut:1999}.
The stability diagram of EC rolls changes qualitatively for $\omega^\prime > \omega^\prime_{C_2}$.
For frequencies $\omega^\prime \gtrsim \omega^\prime_{C_2}$ the abnormal rolls bifurcating at $\varepsilon = \varepsilon_{AR}$ loose stability against the bimodal instability as already discussed for $\omega^\prime  < \omega^\prime_{C_2}$, while for larger $\omega^\prime$ a long-wavelength skewed varicose instability at $\varepsilon = \varepsilon_{SV}$ takes over.
The main features of the phase diagram shown in Fig.~\ref{fig:STABMBBA} for MBBA are similar to those shown in Fig.~3 in \cite{Plaut:1997} for the nematic Phase~5.
The main difference is that here the primary bifurcation at small $\omega^\prime$ is towards oblique rolls, where $\bm q_c$ and $\bm{\hat x}$ include a finite angle.
%

%
%%%% Amplitude equations %%%%
%
\section{Amplitude equations}
\label{sec:Ampl_Eqs}
Amplitude equations have been proven in general to be a very convenient tool to describe roll patterns and their stability near a convection instability \cite{Cross:1993}.
The main idea is to make use of a separation of length and time scales near $\varepsilon =0$.
By suitable techniques fast variations in space and time are projected out to arrive at equations that describe the physics on the slow length scales of the order $1/(\sqrt{\varepsilon} q_c)$ in space and of the order $\tau_d / \varepsilon$ in time.
As already described in the Introduction, Sec.~\ref{intro}, the complete description of EC in the weakly nonlinear regime needs three coupled amplitude equations for the roll amplitude $A$, a vorticity potential $G$ and a quantity $\varphi$ associated with in-plane rotations of the director field.
Following closely the procedure discussed in \cite{Plaut:1999}, the CAE read as follows:
\begin{subequations}
\label{eq:ampAGPHI}
\begin{equation}
\label{EQAMPA}
\begin{split}
\tau_0 \partial_t A &= \varepsilon (A - i e_1 \partial_x A - e_2 \partial_x^2 A - e_3 \partial_y^2 A) \\
&\quad + r_1 \partial_x^2 A + r_2 \partial_y^2 A - |A|^2 A \\
&\quad - i a_1 |A|^2 \partial_x A - i a_2 A^2 \partial_x A^* - a_3 |A|^2 \partial_x^2 A \\
&\quad - a_4 A^2 \partial_x^2 A^* - a_5 |A|^2 \partial_y^2 A - a_6 A^2 \partial_y^2 A^* \\
&\quad - i s_1 A \partial_y G - s_2 A \partial_x \partial_y G \\
&\quad - i b_1 A \partial_y \varphi - i b_2 \varphi \partial_y A + b_3 \varphi \partial_x \partial_y A + b_4 A \partial_x \partial_y \varphi \\
&\quad - \beta_1 A \varphi^2 + i \beta_2 A \varphi \partial_x \varphi + \beta_3 \varphi^2 \partial_y^2 A \,,
\end{split}
\end{equation}
\begin{equation}
\label{EQAMPG}
\begin{split}
0 &= (\nu_a\partial_x^2 + \nu_b \partial_y^2) G + q_1\partial_x\partial_y |A|^2 \\
&\quad + i q_2 \partial_x (A\partial_x\partial_y A^*-c.c.) + i q_3 \partial_y (A^* \partial_x^2 A-c.c.) \\
&\quad + i q_4\partial_y^2 (A^* \partial_y A - c.c.) + \Gamma_G \partial_y^2 (|A|^2\varphi) \,,
\end{split}
\end{equation}
\begin{equation}
\label{EQAMPPHI}
\begin{split}
\partial_t \varphi &= \sigma_T \varphi + K_3 \partial_x^2 \varphi + K_1 \partial_y^2 \varphi + \Gamma_\varphi |A|^2 \varphi - g_\varphi \varphi^3 \\
&\quad - \gamma_1 (A^* \partial_x \partial_y A + c.c.) + i \gamma_2 (A^* \partial_y A - c.c.) \,.
\end{split}
\end{equation}
\end{subequations}
Here $c.c$ stands for complex conjugate.
Compared to \cite{Plaut:1999} the present formulation of the CAE [Eq.~(\ref{eq:ampAGPHI})] contains additional derivatives with respect to $x$.
These are for instance needed to capture the SV instabilities of rolls, which have been identified before in the Galerkin stability analysis (see Fig.~\ref{fig:STABMBBA}).
The numerical values of the coefficients of Eq.~(\ref{eq:ampAGPHI}) have been calculated for MBBA along the lines explained in \cite{Kaiser:1993,Plaut:1999}; for some selected frequencies they are listed in Appendix~\ref{sec:appCoeff}.
The extensive use of {\it Mathematica} was again crucial to perform the tedious but in principle straightforward manipulations.
Our CAE are considered to be a kind of minimal description of planar EC in the weakly nonlinear regime with respect to the stability of the rolls and the patterns that develop in the ensuing unstable regimes.
It is certainly worthwhile and clarifying to give a more detailed interpretation of the fields $A$, $G$, and $\varphi$ appearing in the CAE.
The complex pattern amplitude $A(\bm x, t)$ describes the spatial variations of the rolls in the horizontal plane together with their dynamics.
By multiplying $A$ with $\exp (i q_c x)$ the original fast scale variations are restored.
The resulting quantity is proportional to the $z$ average $\langle n_z(\bm x, t) \rangle$ of the out-of-plane component $n_z(\bm x, z, t)$ of the director field, i.e., 
\begin{eqnarray}
\label{eq:shadow}
\langle n_z(\bm x, t) \rangle \propto  \textrm{Re}[ A(\bm x, t) \exp (i q_c x)] \,. 
\end{eqnarray}
which governs according to \cite{Trainoff:2002,Pesch:2013} in a first approximation the standard shadowgraph pictures of the patterns recorded with the geometry O1 discussed in Sec.~\ref{sec:experim}.
Using the general decomposition $A(x,y,t) = |A| \exp[i \psi(x,y,t)]$ it is obvious that a finite value of $(\partial_x,\partial_y)\psi$ describes spatial modulations of the wave vector $\bm q_c =(q_c,0)$ of an ideal roll pattern, whereas spatial variations of $|A|$ are directly reflected in amplitude modulations.
In this context a zero point $(x_0,y_0)$ of $A(x,y,t)$ corresponds to a dislocation in the striped roll pattern when the circulation of $(\partial_x,\partial_y)\psi$ around $(x_0,y_0)$, the so-called topological charge of the point defect, has the value $\pm 2 \pi$.
The importance of point defects in the EC pattern dynamics has been emphasized in earlier studies (see, e.g., \cite{Bodenschatz2:1988,Kaiser:1993}).
The amplitude $A$ is coupled to the vorticity potential $G(x,y,t)$, which is not directly accessible in experiments.
The point is that any distortion of the perfect roll pattern with wave vector $\bm q_c$, i.e., spatial variations in $A$, leads to a toroidal flow field of the form $\bm V(x, y, z, t) = [ z^2 -(d/2)^2 ] \{ \partial_y, -\partial_x, 0 \} G(x,y,t)$ and thus to a vertical vorticity field $(\textrm{rot} \bm V)_z \propto (\partial_{xx} + \partial_{yy}) G$.
To derive the equation for the in-plane rotation of $\bm n$ we had to project, in line with \cite{Plaut:1999}, the director equations on the leading mode $n_y(x,y,z, t) = \varphi(x,y,t) \cos(\pi z/d)$.
To visualize the dynamics of $\varphi$ in space and time, which is barely reflected in standard shadowgraphy, one needs the special optical setup O2 discussed in Sec.~\ref{sec:experim}.
As a first test of the CAE, we recalculate the stability diagram of rolls near onset.
Normal rolls are described by the constant solutions $A = A_0 =\sqrt{\varepsilon}$ and $\varphi =0$.
It is easy to see that at $\varepsilon = \varepsilon_{AR} = |\sigma_T|/\Gamma_{\varphi}$ this solution becomes unstable against a $\varphi$ perturbation.
Thus the normal rolls are replaced by the abnormal roll solutions $A= A_0$, $\varphi = \varphi_0$ of Eq.~(\ref{eq:ampAGPHI}) which read as follows:
\begin{equation}
\label{EQAPHI0}
A_0 = \sqrt{\frac{\beta_1|\sigma_T| - g_\varphi \varepsilon}{\beta_1 \Gamma_\varphi - g_\varphi}} \,, \;
\varphi_0 = \pm \sqrt{\frac{\Gamma_\varphi \varepsilon -
|\sigma_T|}{\beta_1 \Gamma_\varphi - g_\varphi}} \,, \;
\varepsilon \geq \varepsilon_{AR} \,.
\end{equation}
Since the amplitudes $A_0$ and $\varphi_0$ are constant we have $G \equiv 0$. 
To analyze the stability of these solutions against long-wavelength instabilities with wave vector $\bm s = (s_x,s_y)$ we insert the ansatz $A=A_0+\delta A$, $\varphi=\varphi_0+\delta \varphi$, and $G=\delta G$ with 
\begin{eqnarray}
\label{STABANS}
&&\delta A = (a_1 e^{i \mathbf{s} \cdot \mathbf{r}} 
+ a_2 e^{-i \mathbf{s} \cdot \mathbf{r}}) e^{\sigma t} \,, \;
\delta \varphi = \varphi_1 (e^{i \mathbf{s} \cdot \mathbf{r}} 
+ e^{-i \mathbf{s}\cdot\mathbf{r}}) e^{\sigma t} \,,
\nonumber \\
&&\delta G = g_1 (e^{i \mathbf{s}\cdot\mathbf{r}} 
+ e^{-i \mathbf{s}\cdot\mathbf{r}}) 
\end{eqnarray}
into the CAE [Eq.~(\ref{eq:ampAGPHI})] and keep only the linear terms $\delta 
A$, $\delta G$, and $\delta \varphi$.
It is easy to see that one arrives at a $3 \times 3$-eigenvalue problem in terms of the coefficients $a_1$, $a_2$, and $\varphi_1$ in Eq.~(\ref{STABANS}) which yields the ``nonlinear growth rate'' $\sigma =\sigma(\mathbf{s})$.
Obviously $\sigma >0$ indicates an instability of the constant solutions $A_0$, $\varphi_0$ of Eq.~(\ref{EQAPHI0}).
Note that $\varepsilon$ appears directly as a parameter in Eq.~(\ref{EQAMPA}) for the roll amplitude $A$, whereas $\omega$ comes in via the frequency dependence of the coefficients (see Table~\ref{tab:coeff}).
%

%%% Figure 3
\begin{figure}[ht]
\centering
\includegraphics[width=0.95\linewidth]{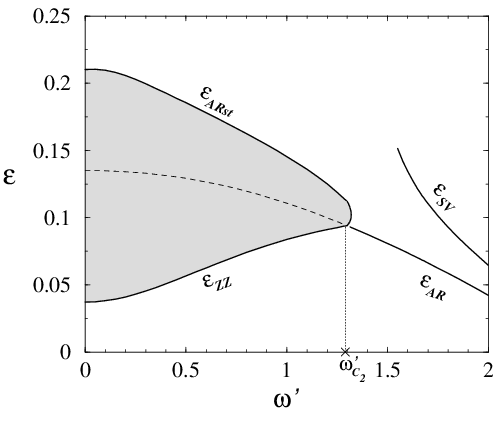}
\caption{\label{fig:STABAMP}
Stability diagram for normal rolls in MBBA in the $\omega^\prime - \varepsilon$ plane in the framework of the CAE.}
\end{figure}
Comparison of the resulting approximate CAE stability diagram in Fig.~\ref{fig:STABAMP} and the exact Galerkin one in Fig.~\ref{fig:STABMBBA} shows that they fit each other fairly well.
This applies in particular to the transition line $\varepsilon_{AR}(\omega^\prime)$ to abnormal rolls and to the characteristic gray-shaded instability regime of the rolls for not too large $\omega^\prime$.
A closer look at the CAE stability analysis shows that the ZZ instability exists only if the particular parameter combination $\delta \Gamma = s_1 \Gamma_G/\nu_b - b_1 \Gamma_{\varphi}/{|\sigma_T|}$ of the $\omega$-dependent coefficients of Eq.~({\ref{eq:ampAGPHI}) is positive.
Thus the condition $\delta \Gamma =0$ was used to determine $\omega^\prime_{C_2} \approx 1.28$ within the CAE approximation, which is slightly larger than the Galerkin value $\omega^\prime_{C_2} =1.15$ given in Sec.~\ref{sec:stabdiag}.
The main discrepancies between the exact Galerkin and the CAE results are observed for $\omega^\prime > \omega^\prime_{C_2}$.
In the Galerkin stability diagram (Fig.~\ref{fig:STABMBBA}) the SV destabilization line $\varepsilon_{SV}$ starts bending up at about $\omega^\prime =1.4$ and would further increase when decreasing $\omega^\prime$.
But this has no relevance for the stability diagram since the bimodal instability line $\varepsilon_{BV}$ takes over.
The latter mechanism is not covered by the CAE and thus in Fig.~\ref{fig:STABAMP} only the strongly rising $\varepsilon_{SV}$ line has survived.
Such quantitative deviations mainly at larger $\varepsilon$ are expected to become smaller by keeping additional gradient terms and to go beyond the cubic terms in the CAE.
But striving for quantitative agreements for larger $\varepsilon$ is in general outside the scope of amplitude equations as a means to explore pattern forming systems.
%

%% Scaling of the CAE %%
%
In the following Sec.~\ref{sec:simul} numerical simulations of the CAE 
[Eq.~(\ref{eq:ampAGPHI})] are compared with corresponding experimental results.
Here the lengths ($x,y$) are measured in units of $d/\pi$ and the time $t$ in units of $\tau_d$ (see Appendix~\ref{sec:appMBBA}).
For simulations ``Ginzburg-Landau (GL) rescalings'' of the CAE are convenient.
They depend then on the dimensionless variables $X,Y$ in space, on $T$ in time, and on the rescaled amplitude $A^\prime$, which are defined as follows:
\begin{equation}
\label{eq:rescal}
A = \sqrt{\varepsilon} A^\prime \,, \;
x = \sqrt{r_1/\varepsilon} X \,, \;
y = \sqrt{r_2/\varepsilon} Y \,, \;
t = (\tau_0 /\varepsilon) T \,. 
\end{equation}
The so-called GL correlation lengths $\sqrt{r_1} > \sqrt{r_2}$ and the GL correlation time $\tau_0$ depend on $\omega^\prime$ and are given in Table~\ref{tab:coeff} for the frequencies discussed in this paper.
For $\omega^\prime =1$, $\varepsilon =0.1$ and $d =25$~$\mu$m we see for instance that the critical wavelength $\lambda_c(\omega^\prime) = (2\pi/q_c(\omega^\prime) (d/\pi)$ corresponds to $\Delta X \approx 0.1$ and a time interval of $1$~s to $\Delta T \approx 1.1$.
%

%
%%% Simulations and comparison with experiments
%
\section{Simulations and comparison with experiments}
\label{sec:simul}
In the standard experiments $\varepsilon$ is slowly increased at fixed $\omega$ (for more details, see Sec.~\ref{sec:experim}).
Thus one identifies the various destabilization mechanisms and gets insight into the pattern that develops after an instability.
To investigate them in theory we consider the use of CAE to be at the moment the only feasible way.
The rescaled CAE (see the previous section) are integrated on a suitably chosen square with side lengths $L_X = L_Y$ in the $X,Y$ plane.
Since we use periodic boundary conditions, the calculations are performed in Fourier space; the nonlinearities are treated by pseudo-spectral methods, whereby extensive use of FFT (fast Fourier transformation) methods is made.
We use typically a $256 \times 256$ grid with the same grid size $\Delta Q$ in the $X,Y$ directions determined by the condition $L_X = L_Y = 2\pi / \Delta Q$.
This square corresponds in the physical $x,y$ plane [see Eq.~(\ref{eq:rescal})] to a rectangle of side lengths $l_x$, $l_y$ with $l_x/l_y =\sqrt{r_2/r_1}$, which are given in units of $\lambda_c$ below.
Note that according to Table~\ref{tab:coeff} $q_c$, $r_1$, $r_2$,  depend on $\omega^\prime$.
The results of the simulations are visualized as gray scale pictures of $|A(X,Y)|$ and $\varphi(X,Y)$, where maxima are mapped to ``white'' and minima to ``black''.
To compare our simulations directly with experiments we have in general to return to the physical spatial variables $x,y$ with use of Eq.~(\ref{eq:rescal}).
In the following we will in particular visualize $\textrm{Re}[A(\bm x) \exp(i q_c x)]$ [Eq.~(\ref{eq:shadow})].
The resulting, so-called fast-scale-$A$ ({\it fs-$A$}), pictures represent well experimental shadowgraph pictures, recorded with the optical setup O1.
In contrast, $\varphi(x,y)$ is directly accessible in experiments via the setup O2.
To make the {\it fs-$A$} structures in the $x,y$ plane better visible, we cut out smaller rectangles of side lengths $L^c_X < L_X$, $L^c_Y < L_Y$ from the original pictures of $|A|(X,Y)$, $\varphi(X,Y)$, where we choose $L^c_X/L^c_Y =\sqrt{r_2/r_1}$.
These cutouts, which are marked by rectangular frames in the figures below, transform thus according to Eq.~(\ref{eq:rescal}) to squares in the $x,y$ plane with side lengths $l^c_x = l^c_y$.
Counting the number $n_r$ of the roll pairs in the {\it fs-$A$} pictures, i.e., the number of white (or black) stripes, allows us to immediately assess the physical size of a cutout via $l^c_x = l^c_y = n_r \lambda_c$.
A quantitative agreement between theory and experiment is not to be expected, as apart from the complicated optics involved in shadowgraphy \cite{Trainoff:2002,Pesch:2013} the experimental pictures are typically digitally processed to enhance their contrast.
%

%
%%%
\subsection{Frequency regime $\omega^\prime < \omega^\prime_{C_2}$}
At first we will concentrate on frequencies $\omega^\prime < \omega^\prime_{C_2}$, where the normal rolls become ZZ unstable at  $\varepsilon =\varepsilon_{ZZ}$ (see Figs.~\ref{fig:STABMBBA} and \ref{fig:STABAMP}).
In our CAE simulations for this regime the ZZ-instability leads in general for $\varepsilon \gtrsim \varepsilon_{ZZ}$ to slow undulations of the rolls along their axes as predicted in Eq.~(\ref{eq:longwav}).
At larger $\varepsilon$ the undulations develop then into the so-called zigzag patterns in the form of alternating domains of oblique rolls with wave vectors $\bm q = (q_c, \pm p)$, which are separated by sharp walls.
This general scenario has been already observed in fairly old experiments \cite{Ribotta:1986,Joets:1991,Nasuno:1992,Rudroff:1998}.
%

%%% Figure 4
\begin{figure}[ht]
\centering
\includegraphics[width=0.45\linewidth]{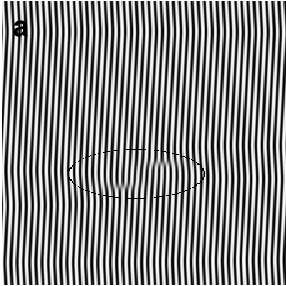}
\hspace{0.02\linewidth}
\includegraphics[width=0.45\linewidth]{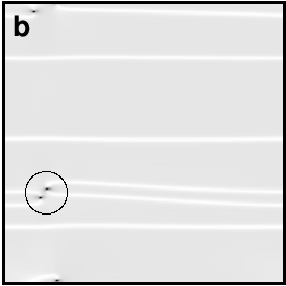}
\caption{\label{ZZROLL2}
Transient zigzag pattern clearly visible in a {\it fs-$A$} picture ($Re[A(x,y) \exp(i q_c x)]$) (a) and the modulus of the roll amplitude $|A(X,Y)|$ (b) for $\omega^\prime =1$ and $\varepsilon =0.1$.}
\end{figure}
To clarify the details of the ZZ destabilization process of rolls we discuss in the following representative simulations for the frequency $\omega^\prime =1 < \omega^\prime_{C_2}$ on a square with side lengths $(l_x,l_y) =(250, 115) \lambda_c$ in physical space.
For $\varepsilon < \varepsilon_{ZZ} (\simeq 0.084)$ we arrive indeed at stable normal roll patterns (not shown) when starting the simulations of the CAE with random initial conditions.
For $\varepsilon$ slightly above $\varepsilon_{ZZ}$ the regular smooth undulations of the rolls along their axis are clearly visible in the {\it fs-$A$} picture in Fig.~\ref{ZZROLL2}(a).
In Fig.~\ref{ZZROLL2}(b) we show $|A(X,Y)|$.
Here the white horizontal lines indicate maxima of $|A|$.
They correspond to lines of $\varphi =0$, which separate horizontal domains with twist field $\pm \varphi$ [see Eq.~(\ref{EQAPHI0})].
During the temporal evolution of the undulated roll structure we observe a complicated transient defect dynamics.
Here point defects in the form of zeros of $|A(X,Y)|$ [as already discussed in more detail after Eq.~(\ref{eq:shadow})] play an important role.
In the experimental roll pattern zeros of $A$ appear as dislocations.
This is evident by concentrating on the two dislocations inside the ellipse with the ratio $\sqrt{r_1/r_2}$ of the major and minor axis in the lower part of Fig.~\ref{ZZROLL2}(a), which corresponds to the circular region in Fig.~\ref{ZZROLL2}(b).
At later times the two point defects in the circular region with opposite topological charges will annihilate each other and the two adjacent white lines will coalesce, such that eventually a regular undulated roll pattern will cover the whole integration domain.
%

%%% Figure 5
\begin{figure}[ht]
\centering 
\includegraphics[width=0.45\linewidth]{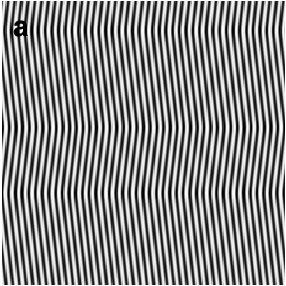} 
\hspace{0.02\linewidth}
\includegraphics[width=0.45\linewidth]{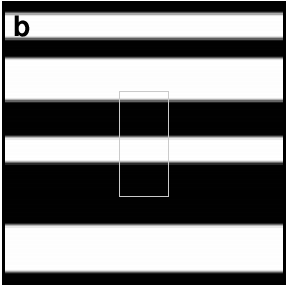}
\caption{\label{ZZROLL1}
Quasistationary {\it fs-$A$} zigzag pattern (a) and the corresponding twist field $\varphi$ (b) for $\omega^\prime =1$ and $\varepsilon =0.12$.}
\end{figure}
%

%%% Figure 6
\begin{figure}[ht]
\centering
\includegraphics[width=0.45\linewidth]{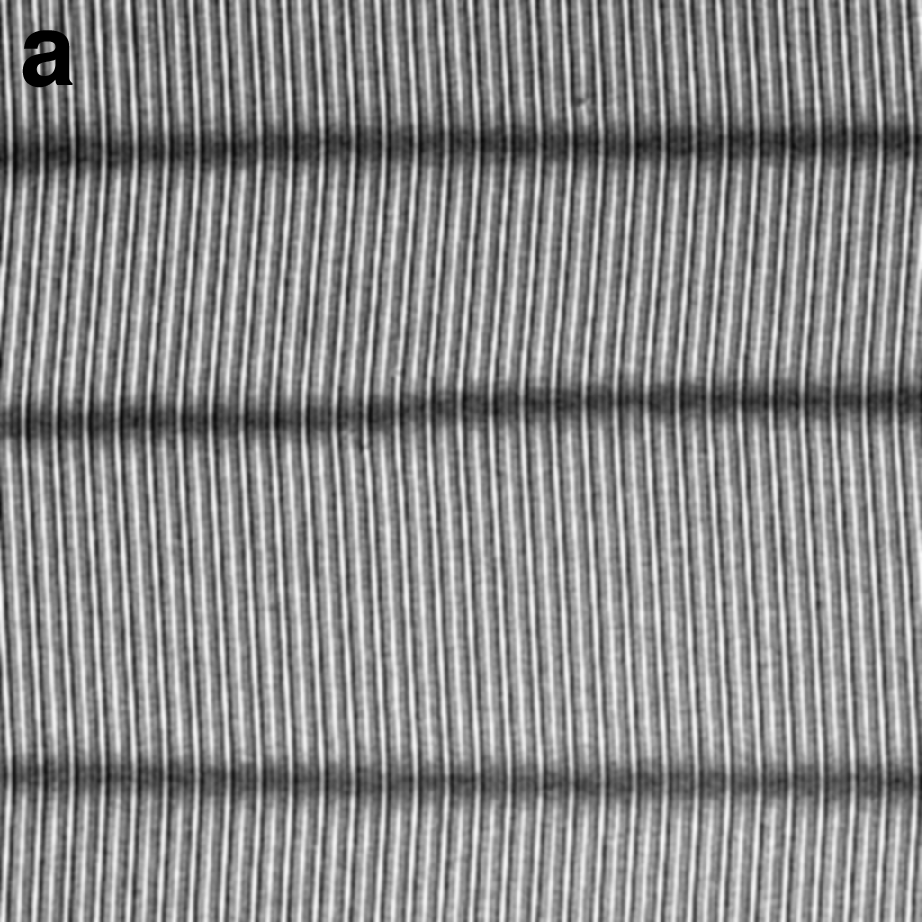}
\hspace{0.02\linewidth}
\includegraphics[width=0.45\linewidth]{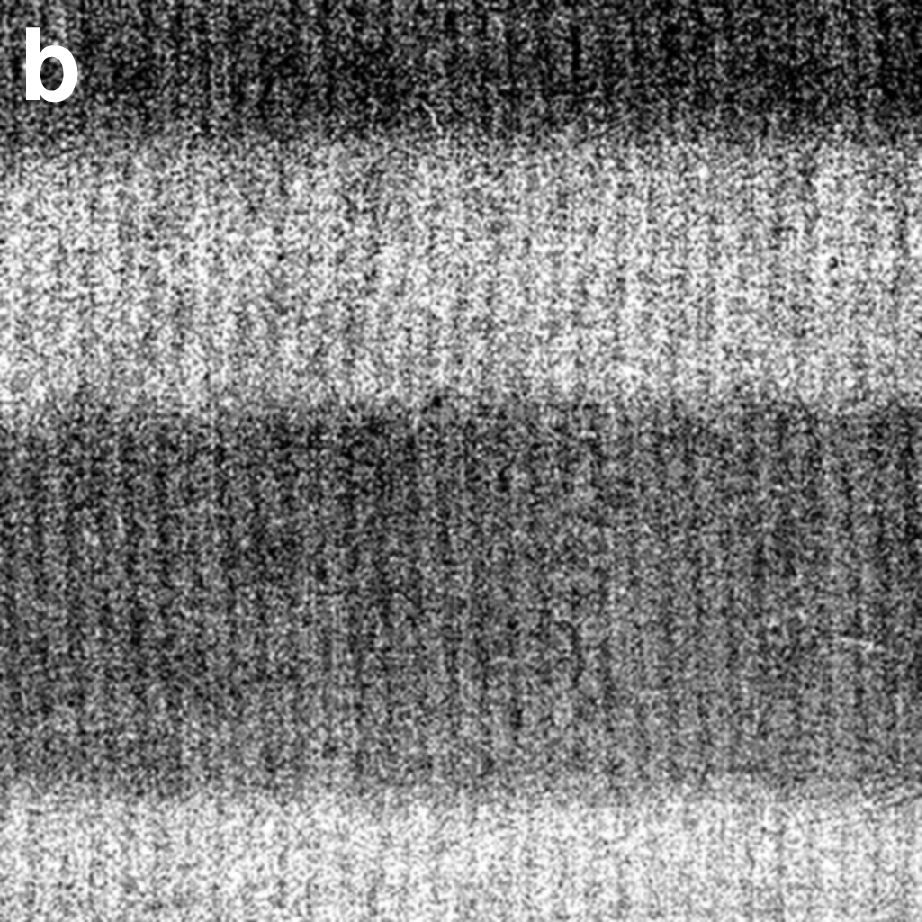}
\caption{\label{fig:wtq1_eps0_12}
Experimental snapshots for $\omega^\prime =1$ and $\varepsilon =0.12$, above threshold of zigzag instability: optical geometry O1 (a) and optical geometry O2 (b).
The image size is $800$~$\mu$m~$\times \, 800$~$\mu$m.}
\end{figure}
For $\varepsilon > \varepsilon_{AR} \simeq 0.11$ the undulations develop into zigzag patterns as shown in Fig.~\ref{ZZROLL1}(a) for $\varepsilon =0.12$, where {\it zig} and {\it zag} domains alternate along the $y$ direction.
The final pattern, which has developed after about $T =2 \times 10^4$ ($\sim 4.2$~h in physical units) is stationary.
Note that this process takes typically a much longer time in our deterministic simulations than in experiments in the presence of imperfections of the cell and thermal noise.
Inspection of the corresponding {\it twist} amplitude $\varphi(X,Y)$ [Fig.~\ref{ZZROLL1}(b)] demonstrates that the zig and zag regions correspond indeed to the white (black) regions with $\varphi >0$ ($\varphi <0$); the framed rectangular cutout in Fig.~\ref{ZZROLL1}(b) has been used to construct the {\it fs-$A$} pattern in Fig.~\ref{ZZROLL1}(a).
The results of the simulations are very well confirmed by the experimental quasistationary zigzag pattern shown for comparison in Fig.~\ref{fig:wtq1_eps0_12}.
Note that similar pictures of the zig-zag instability of normal rolls have been observed in the nematic Phase~5 \cite{Ribotta:1986}.
%

%
%%%
\subsection{Frequency regime $\omega^\prime > \omega^\prime_{C_2}$}
Our main focus in this paper is on the parameter regime $\omega^\prime > \omega^\prime_{C_2}$ of the CAE-phase diagram in Fig.~\ref{fig:STABAMP} which has so far attracted less interest in systematic experimental studies.
This regime deserves certainly a more detailed investigation to understand a number of interesting, complex structures superseded to the EC roll patterns, like regular chains of defects, line defects, and defect-chaotic states, which have been described previously in the literature \cite{Joets:1991,Nasuno:1992,Oikawa:2004}.
According to Fig.~\ref{fig:STABAMP} the primary bifurcation of the normal rolls at $\varepsilon =\varepsilon_{AR}$ leads to the abnormal rolls, which are characterized by a homogeneous in-plane rotation ($\pm \varphi$) of the director.
As long as $\varepsilon < \varepsilon_{SV}$, where the abnormal rolls become unstable against the skewed varicose instability (see Fig.~\ref{fig:STABAMP}), the simulations show clearly, that the abnormal rolls for $\varepsilon > \varepsilon_{AR}$ have the strong tendency to develop in general striped domains with alternating angles $\pm \varphi$, which are separated by straight walls.
It is remarkable that those domains have a strong preference for a horizontal orientation (perpendicular to the roll axes).
This behavior is documented in a representative example for $\omega^\prime =1.6$ at $\varepsilon =0.1$ (above $\varepsilon_{AR} \simeq 0.074$) in Fig.~\ref{ARROLLH01}, where the {\it fs-$A$} pattern [panel (a)] corresponds to the framed cutout of the $\varphi$ field [panel (b)]; the full integration domain corresponds to $(l_x,l_y) =(220, 150) \lambda_c$.
Note that the two horizontal {\it abnormal} roll domains are not reflected in the roll pattern.
The pictures shown in Fig.~\ref{ARROLLH01} are in fact not stationary, since the horizontal walls drift globally with a constant velocity upward in $y$ direction without any influence on the shape of the domains.
As a typical value for the drift velocity we find $v_D \approx 0.055$ ($\approx 7$~$\mu$m/s in physical units). 
Note that by symmetry reasons the upward or downward drifts are equally possible in our system; the actual direction is selected by the initial conditions.
%

%%% Figure 7
\begin{figure}[ht]
\centering 
\includegraphics[width=0.45\linewidth]{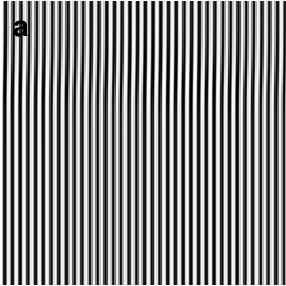} 
\hspace{0.02\linewidth}
\includegraphics[width=0.45\linewidth]{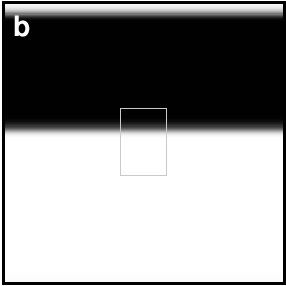}
\caption{\label{ARROLLH01}
Abnormal fs-$A$ roll pattern (a) and the corresponding twist field $\varphi$ (b) for $\omega^\prime =1.6$ and $\varepsilon =0.1$.}
\end{figure}
Experimental pictures of horizontal abnormal roll domains taken in the two optical geometries are shown in Fig.~\ref{fig:wtq1_6_eps0_11}.
Since the contributions to the optical image from out-of-plane ($n_z$) and in-plane ($n_x, n_y$) director distortions cannot be completely separated the horizontal grain boundary is also visible in Fig.~\ref{fig:wtq1_6_eps0_11}(a) in contrast to Fig.~\ref{ARROLLH01}(a).
The horizontal grain boundary between abnormal rolls moves here upward along the roll axis with a drift velocity of about $8$~$\mu$m/s.
%

%%% Figure 8
\begin{figure}[ht]
\centering
\includegraphics[width=0.45\linewidth]{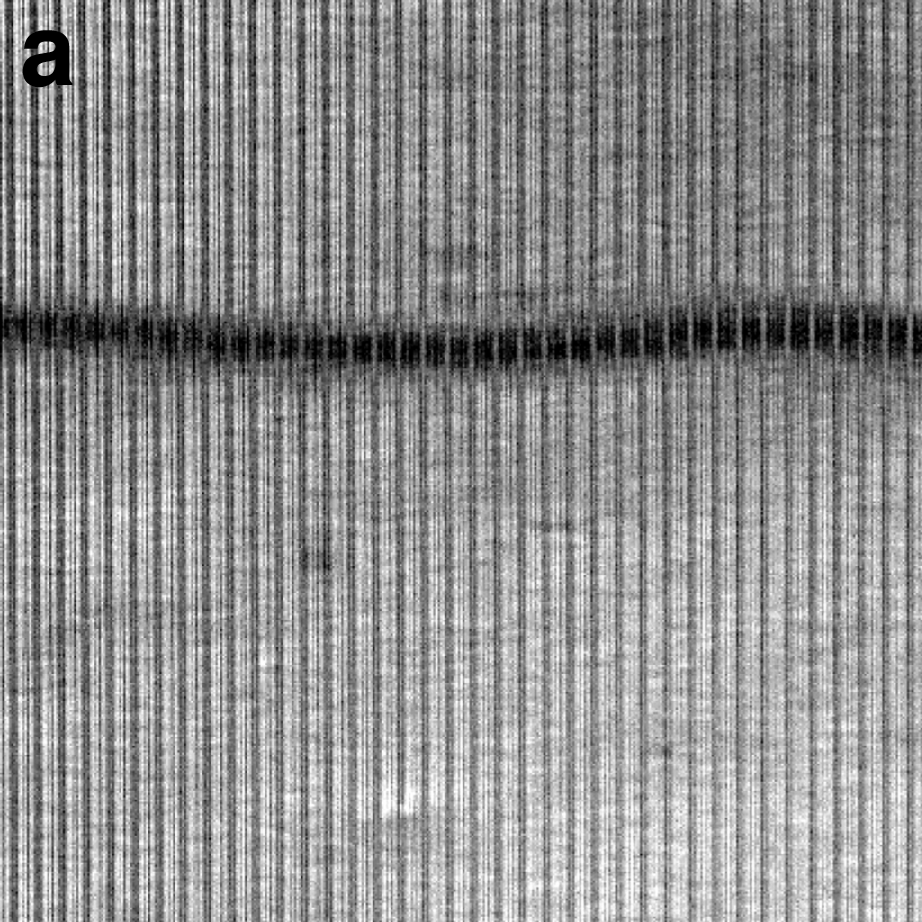}
\hspace{0.02\linewidth}
\includegraphics[width=0.45\linewidth]{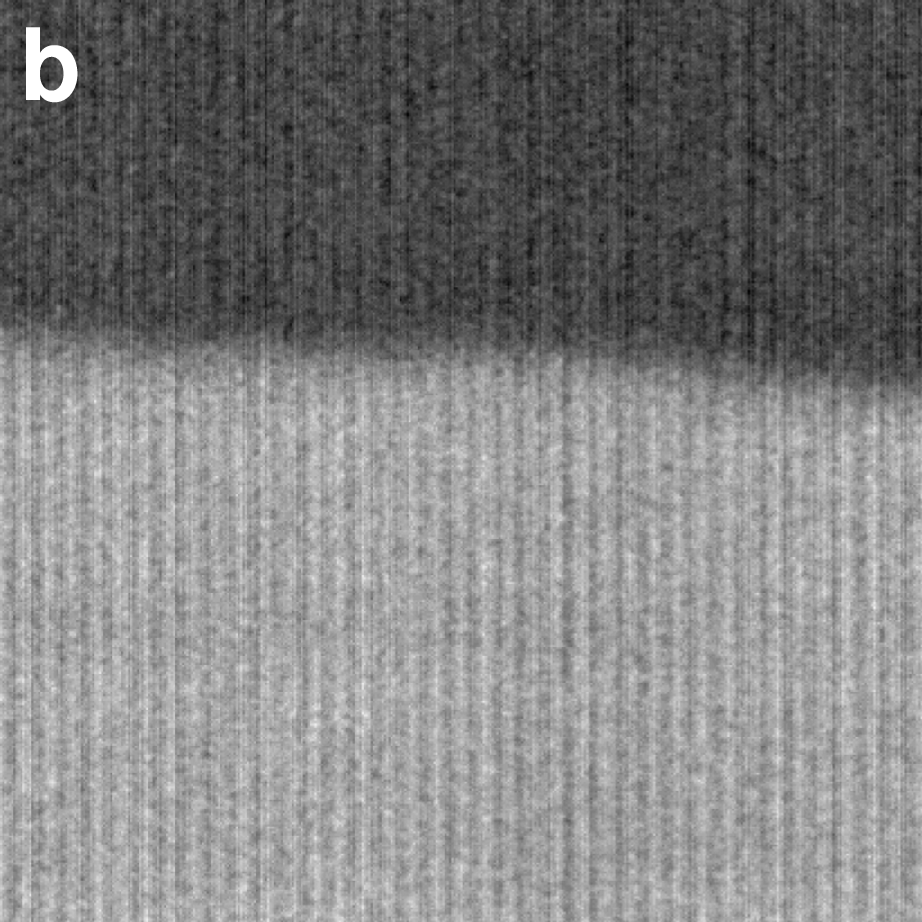}
\caption{\label{fig:wtq1_6_eps0_11}
Experimental snapshots for $\omega^\prime =1.6$ and $\varepsilon = 0.11$, above threshold of abnormal rolls: optical geometry O1 (a) and optical geometry O2 (b).
The image size is $800$~$\mu$m~$\times \, 800$~$\mu$m.}
\end{figure}
With increasing $\varepsilon$ the walls between the $\pm \varphi$ domains become steeper and the drift velocities smaller [e.g., $v_D =0.013$ ($1.6$~$\mu$m/s in physical units) at $\varepsilon=0.13$].
At about $\varepsilon =0.135$ the patterns seem to become stationary.
Above $\varepsilon =\varepsilon_{SV} \approx 0.142$ for $\omega^\prime =1.6$ the abnormal roll patterns like the one shown in Fig.~\ref{ARROLLH01} start to become unstable.
They show a weakly chaotic dynamics, which is characterized by a permanent generation and annihilation of domain walls and dislocations. 
Figure~\ref{DEFROLL2}(a) presents a characteristic {\it fs-$A$} picture  which corresponds to the framed cutout in the pictures of the twist field $\varphi$ and the modulus of the roll amplitude $|A|$ shown in Figs.~\ref{DEFROLL2}(b) and \ref{DEFROLL2}(c), respectively.
The simulations are well confirmed by corresponding experimental snapshots taken in the two optical geometries O1, O2, which are shown in Fig.~\ref{fig:wtq1_6_eps0_2}.
In fact they look also very similar to the experimental Fig.~8(d) in \cite{Nasuno:1992}, which was, however, taken at a much higher value of the control parameter $\varepsilon$.
%

%%% Figure 9
\begin{figure*}[ht]
\centering 
\includegraphics[width=0.225\linewidth]{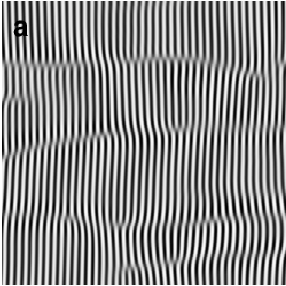} 
\hspace{0.01\linewidth}
\includegraphics[width=0.225\linewidth]{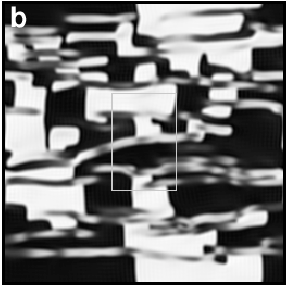} 
\hspace{0.01\linewidth}
\includegraphics[width=0.225\linewidth]{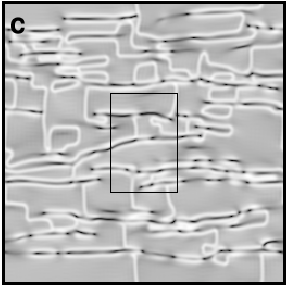}
\caption{\label{DEFROLL2}
Defect chaotic {\it fs-$A$} patterns (a), the corresponding twist field $\varphi$ (b), and the modulus of the roll amplitude $|A|$ (c) for $\omega^\prime =1.6$ and $\varepsilon =0.19$.}
\end{figure*}
%

%%% Figure 10
\begin{figure}[ht]
\centering
\includegraphics[width=0.45\linewidth]{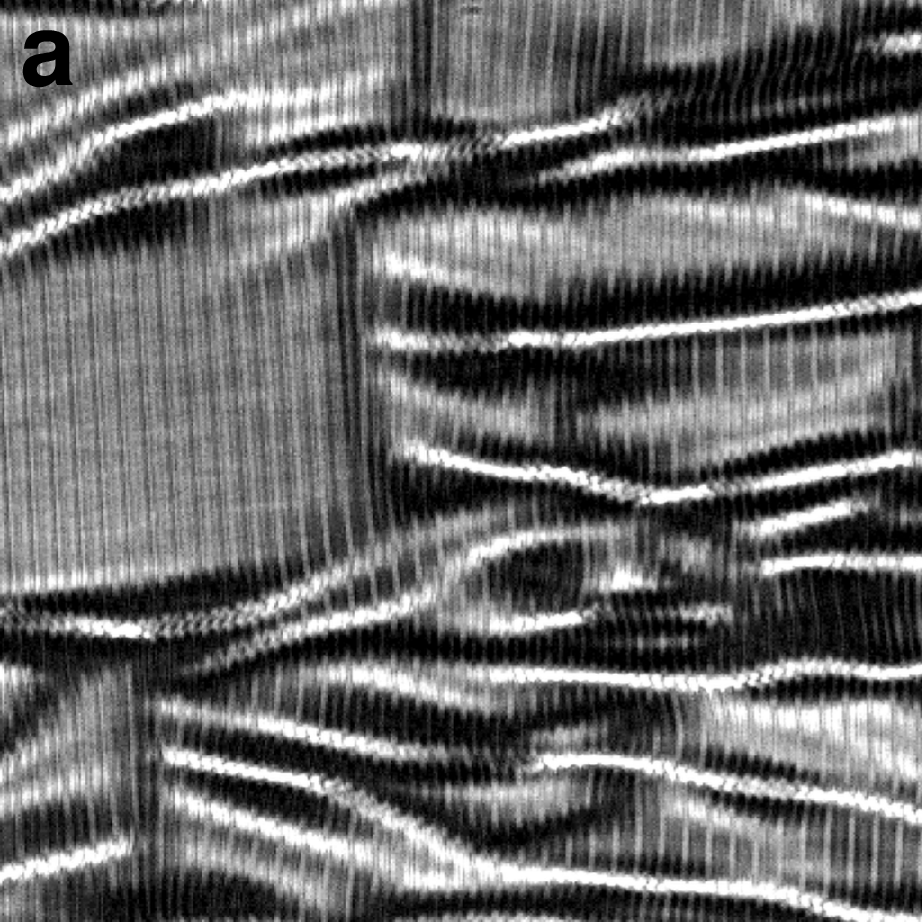}
\hspace{0.02\linewidth}
\includegraphics[width=0.45\linewidth]{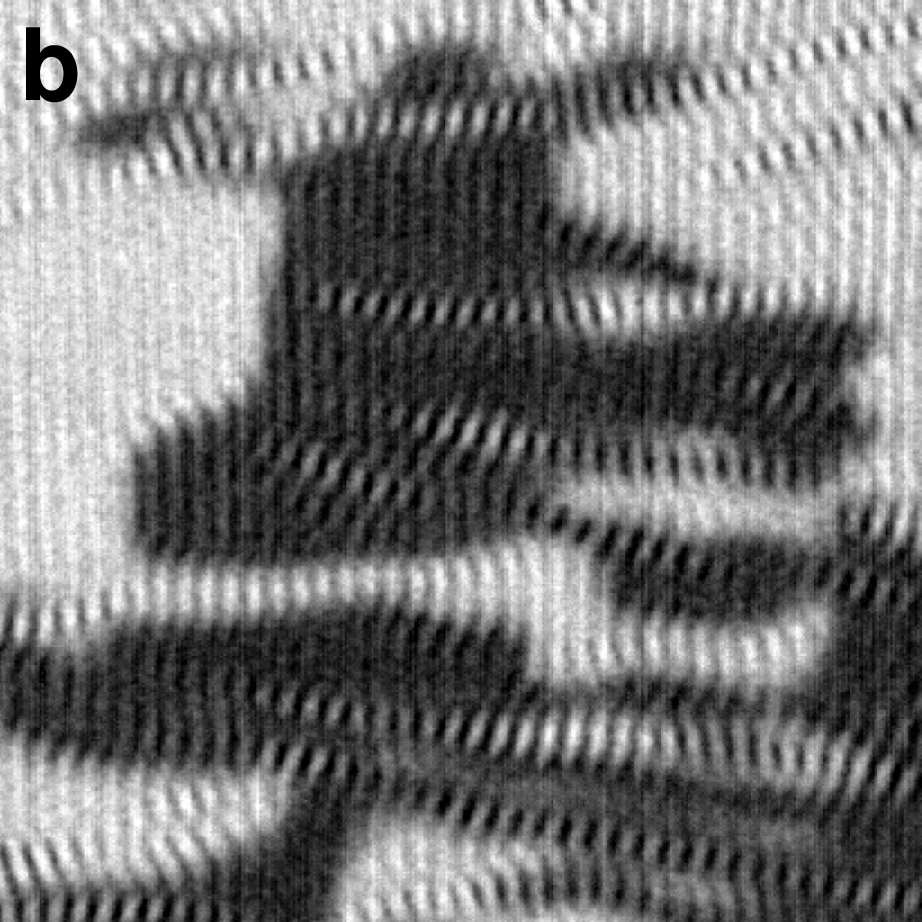}
\caption{\label{fig:wtq1_6_eps0_2}
Experimental snapshots of chaotic defect patterns for $\omega^\prime =1.6$ and $\varepsilon =0.2$: optical geometry O1 (a) and optical geometry O2 (b).
The image size is $800$~$\mu$m~$\times \, 800$~$\mu$m.}
\end{figure}
In the following we study the patterns for $\varepsilon > \varepsilon_{AR}$ at larger $\omega^\prime$.
It turns out that at $\omega^\prime =\omega^\prime_{HV} \approx 1.67$ the general morphology of the patterns changes qualitatively.
As described before the $\varphi$ patterns are characterized in the interval $\omega^\prime_{C_2} < \omega^\prime < \omega^\prime_{HV}$ by {\it horizontal domain walls} running perpendicular to the rolls and separating horizontal stationary $\varphi$ domains, where $\varphi$ alternates between the symmetry degenerate states ($\pm \varphi$).
In distinct contrast we obtain for $\omega^\prime > \omega^\prime_{HV}$ and $\varepsilon > \varepsilon_{AR}$ typically $\varphi$ patterns with {\it vertical domain walls} running parallel to the roll axes.
The typical scenarios are discussed in the sequel for $\omega^\prime =1.9$.
%

%%% Figure 11
\begin{figure*}[ht]
\centering
\includegraphics[width=0.225\linewidth]{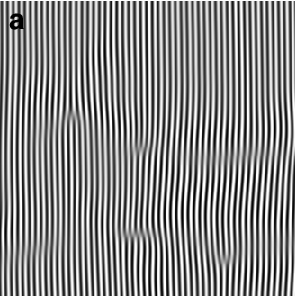}
\hspace{0.01\linewidth}
\includegraphics[width=0.225\linewidth]{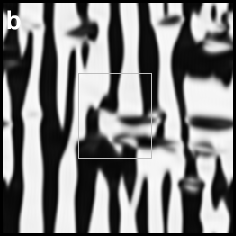}
\hspace{0.01\linewidth}
\includegraphics[width=0.225\linewidth]{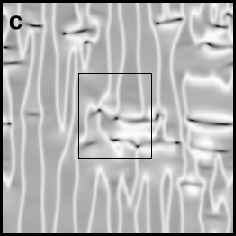}
\\ \vspace*{1cm}
\includegraphics[width=0.225\linewidth]{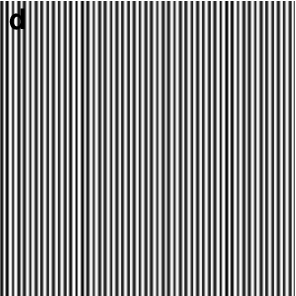}
\hspace{0.01\linewidth}
\includegraphics[width=0.225\linewidth]{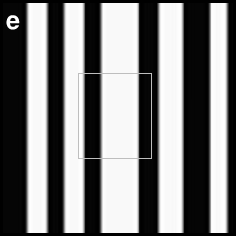}
\caption{\label{ARABSPHIV}
Formation of vertical domain walls for $\omega^\prime =1.9$ and $\varepsilon =0.11$: {\it fs-$A$} picture (a), the corresponding twist field $\varphi$ (b), and modulus of rolls amplitude $|A|$ (c) after  integration time $T =10^4$.
The corresponding pictures, {\it fs-$A$} (d) and $\varphi$ (e) after continuing the simulations by $T =10^4$.}
\end{figure*}
%

%
%%% Figure 12
\begin{figure}[ht]
\centering
\includegraphics[width=0.45\linewidth]{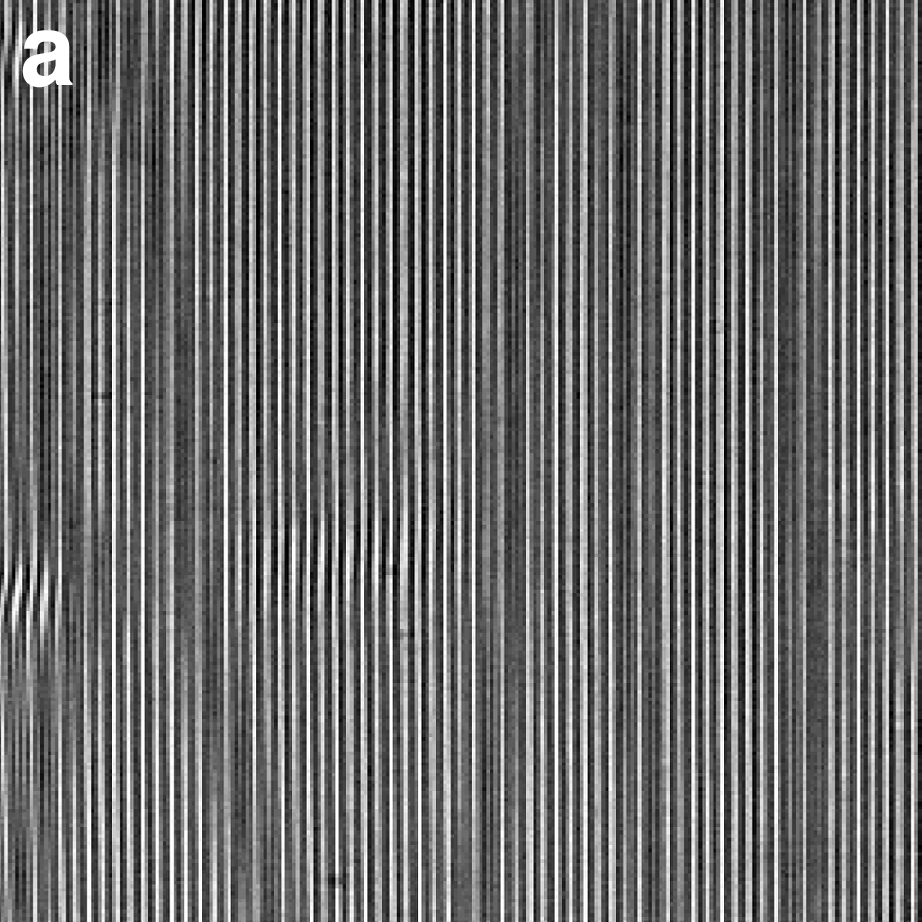}
\hspace{0.02\linewidth}
\includegraphics[width=0.45\linewidth]{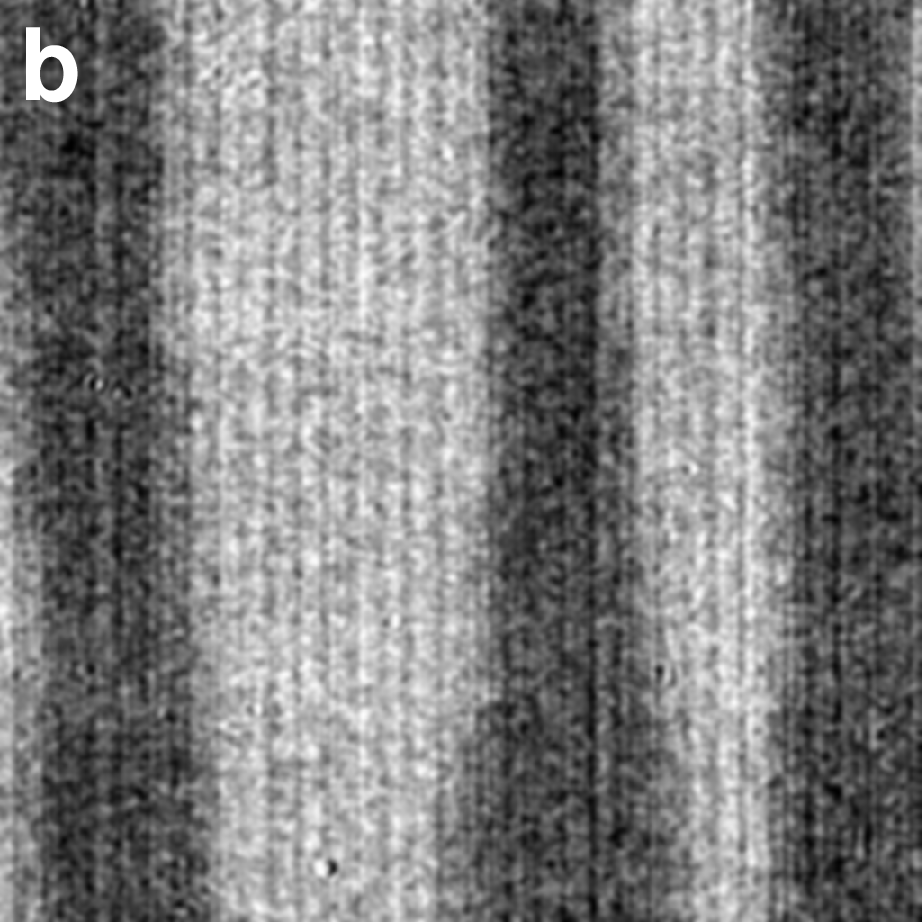}
\caption{\label{fig:wtq1_9_eps0_1}
Experimental snapshots for $\omega^\prime =1.9$ and $\varepsilon = 0.1$: optical geometry O1 (a) and optical geometry O2 (b).
The image size is $500$~$\mu$m~$\times \, 500$~$\mu$m.}
\end{figure}
Figure~\ref{ARABSPHIV}(top) shows for $\varepsilon =0.11 > \varepsilon_{AR} \simeq 0.051$ a {\it fs-$A$} roll picture together with the corresponding pictures of $\varphi$ and of $|A|$, which have developed after $T =10^4$ ($\sim 37$~min in physical units) when starting from random initial conditions.
The simulations cover an area $(l_x,l_y) =(155,135) \lambda_c$.
A tendency to a vertical wall structure is clearly visible in the $|A|$ picture; in addition, the {\it skewed varicose} instability of the rolls ($\varepsilon_{SV} =0.078$) is reflected in dislocations (black points in $|A|$).
Continuing the simulation further by $T =10^4$ the pattern becomes eventually stable against the generation of dislocations leading to the regular final stationary state of vertical $\pm \varphi$ domains separated by the $\varphi =0$ lines (walls) in Fig.~\ref{ARABSPHIV}(bottom).
Typical experimental pictures for this case are shown in Fig.~\ref{fig:wtq1_9_eps0_1} for the two geometries O1, O2.
As demonstrated by simulations shown in Fig.~\ref{PHIWALLCRE} an increase of $\varepsilon$ leads to a destabilization of the vertical domain structure such that their average width gets smaller.
During this process regular distortions develop along a single domain, which cause eventually its splitting as clearly visible when concentrating on the time evolution of the second domains on the left.
The details of this process will be elucidated in Fig.~\ref{AWALLCRE} by following the dynamics of $|A|$ in the small framed part in the lower left corner of Fig.~\ref{PHIWALLCRE}.
Note that the  stripes of maximal $|\varphi|$ (either white or black in Fig.~\ref{PHIWALLCRE}) correspond to minima of $|A|$ (black), while the walls in Fig.~\ref{PHIWALLCRE} ($\varphi =0$) appear as white lines (maxima of $|A|$).
Starting with a $\varphi$ domain between the vertical white lines in Fig.~\ref{AWALLCRE} (picture~0), the skewed varicose instability is responsible for the periodically modulated shear lines (black lines in the $|A|$ field) along the $y$ direction (picture~1).
At their centers a spontaneous generation of dislocation pairs is observed.
Subsequently the defects (black points in the $|A|$ field) move in opposite directions to stop at the vertical walls (picture~2).
Later the process repeats (picture~3), but the single dislocations arriving at 
the walls annihilate with the previous ones.
As a consequence we observe along the center line of the domain a periodic sequence of white closed quasi-annular patches (picture~4) with increased roll amplitude $|A|$.
In the center of these patches the sign of $\varphi$ is opposite to that of the original uniform domain.
Subsequently the center patches contract along the $x$ direction, the horizontal connection are cut (picture~6), and two additional domain walls appear (picture~7).
The whole process takes only about $T =700$ ($\sim 2$~min in physical units).
%

%%% Figure 13
\begin{figure}[ht]
\centering 
\includegraphics[width=0.98\linewidth]{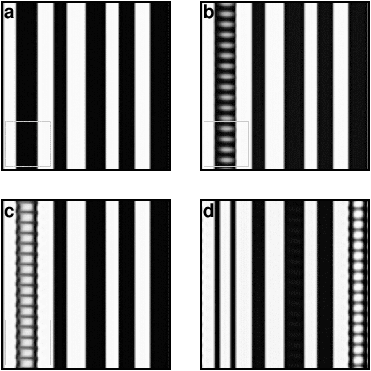} 
\caption{\label{PHIWALLCRE}
Wall dynamics viewed via twist field $\varphi$ during the generation of new walls for $\omega^\prime =1.9$ and $\varepsilon =0.13$. 
The single pictures are shown at times $T =351$ (b), $367$ (c), and $950$ (d) when starting with picture (a).}
\end{figure} 
%

%%% Figure 14
\begin{figure}[ht]
\centering 
\includegraphics[width=0.98\linewidth]{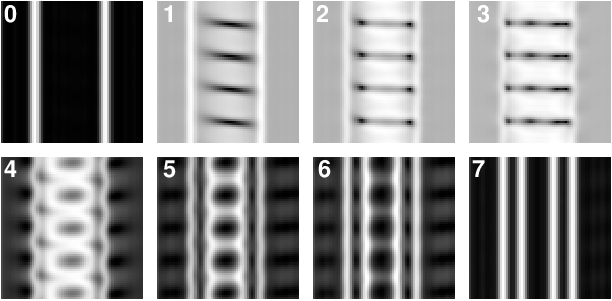} 
\caption{\label{AWALLCRE}
Time sequence of the defect- and wall dynamics in $|A|$ during the generation of a pair of new walls for the parameters of Fig.~\ref{PHIWALLCRE}. 
The pictures are shown for the times $T =340$ (1), $351$ (2), $358$ (3), $367$ (4), $382$ (5), $389$ (6), and $700$ (7) after the first one (0).}
\end{figure}
In essence, it has been demonstrated that the generation of new $\pm \varphi$ domains does not involve the whole pattern, but is rather concentrated to a single wall.
In fact we see the initial state of a new splitting process again at the right-hand side of Fig.~\ref{PHIWALLCRE}(d).
When increasing $\varepsilon$ the domain splitting process continues until only about three rolls belong to a single domain wall.
At $\varepsilon =0.17$ also this structure is destabilized by the SV process. 
At the end we arrive at so-called defect lattices, a typical example of which is shown in Fig.~\ref{DEFLROLL2}.
It is characterized by a regular sequences of vertical $\varphi$ domains which contain periodically repeating walls along the domains [panel (b)].
Each domain [see panel (a)] contains about three rolls which are periodically disturbed by short dislocation lines.
They are related to the short horizontal walls in the $\varphi$ field in panel (b).
%

%%% Figure 15
\begin{figure}[ht]
\centering 
\includegraphics[width=0.45\linewidth]{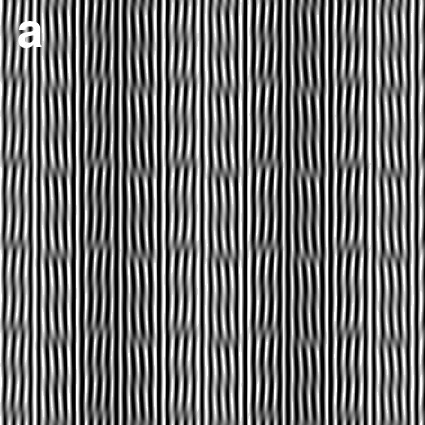}
\hspace{0.02\linewidth}
\includegraphics[width=0.45\linewidth]{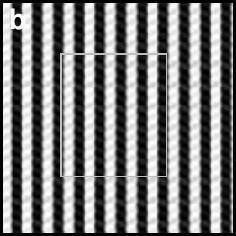}
\caption{\label{DEFLROLL2}
Defect lattice in the roll pattern {\it fs-$A$} (a) and the corresponding twist field $\varphi$ (b) for $\omega^\prime =1.9$ and $\varepsilon =0.19$.}
\end{figure}
%

%%% Figure 16
\begin{figure}[ht]
\centering
\includegraphics[width=0.45\linewidth]{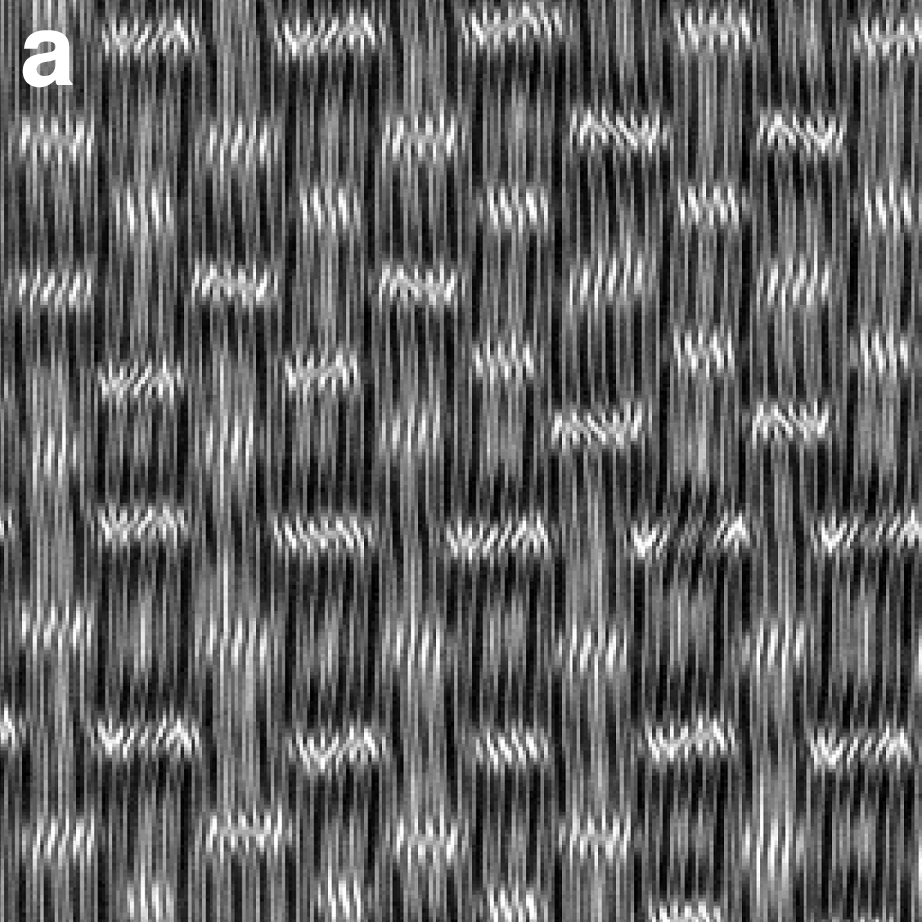}
\hspace{0.02\linewidth}
\includegraphics[width=0.45\linewidth]{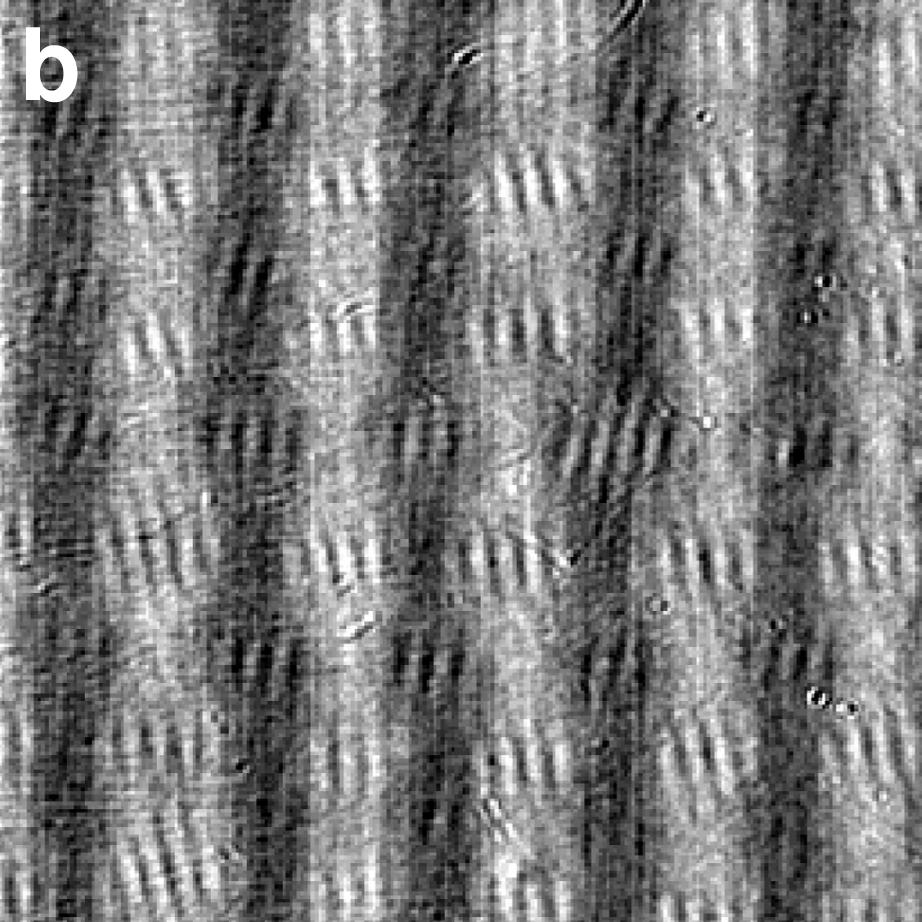}
\caption{\label{fig:wtq1_9_eps0_18}
Experimental snapshots for $\omega^\prime =1.9$ and $\varepsilon = 0.18$: optical geometry O1 (a) and optical geometry O2 (b).
The image size is $500$~$\mu$m~$\times \, 500$~$\mu$m.}
\end{figure}
For comparison we present in Fig.~\ref{fig:wtq1_9_eps0_18} corresponding experimental pictures of a defect lattice taken with the optical setups O1, O2.
As in the simulations the vertical $\varphi$ domains [panel (b)] demonstrate periodic sequences of short horizontal walls.
The corresponding roll patterns show also domains with about three rolls [panel (a)] superseded with short bright horizontal structures.
Note that in the simulations the horizontal structures appear dark since they are caused by regions of reduced amplitude $A$.
In contrast, they appear bright in the experiments since light can pass them without strong deflection.
%

%
%%%% Summary and outlook %%%%
%
\section{Summary and outlook}
\label{sec:summary}
The good agreement between the theory on the basis of the CAE and the experiment proves without doubt the reliability of our theoretical approach.
It is in any case very useful to have visualized not only the convection rolls as standard in EC experiments but also the director configuration by the special optical setup O2, which was rarely done in the past.
The CAE [Eq.~(\ref{eq:ampAGPHI})] look certainly very complicated and contain quite a number of coefficients.
Thus it was crucial to calculate those directly from the basic nemato-hydrodynamic equations, such that our theory does not contain adjustable parameters except the frequency scale (determined by the fit parameter $\tau_q$).
In fact, some qualitative features of experimental EC patterns in planar nematics have been reproduced by numerical simulations of simpler CAE versions.
Their construction is based on symmetry arguments and the coefficients are treated as free parameters.
Furthermore, the vorticity amplitude $G$ [Eq.~(\ref{EQAMPG})], relevant to describe properly the dynamics of defects in the patterns, is not included.
One example of such simplified CAE can be found in \cite{Rossberg2:1998,Rossberg:1998}.
Some of the resulting simulation pictures (see in particular Fig.~4.2 in \cite{Rossberg:1998}) show indeed some superficial similarities to ours shown in Sec.~\ref{sec:simul}.
We doubt strongly, however, that this is sufficient for a reliable validation of these simplified CAE.
For instance, when mapping our general CAE to those used in \cite{Rossberg:1998}, nonrealistic values of the coefficients are needed.
We have to allow for a sign change of $b_1$ in Eq.~(\ref{EQAMPA}) at larger $\omega^\prime$, while in Table~\ref{tab:coeff} this coefficient is really always positive and monotonically increasing as a function of $\omega^\prime$.
Furthermore, we had to increase our ``effective elastic constant'' $K_1$ in Eq.~(\ref{EQAMPPHI}) by a factor of $5 - 10$.
A different set of quite simple CAE, proposed in \cite{Komineas:2003}, is certainly worth mentioning as well.
By an interesting analysis the existence and stability of vertical domains was clearly proven in this approach, which supports the present simulations of our full CAE.
In the near future a more detailed exploration of the CAE solution manifold is planned.
For instance, one would like to reveal the mechanism responsible for the transition from vertical walls to horizontal walls.
It would be also rewarding to construct more general CAE for a combined ac and dc driving voltage to provide a theoretical interpretation of oscillating grid patterns \cite{Batyrshin:2012} recently observed in experiments.
%

%
%%%% Acknowledgments %%%%
%
\begin{acknowledgments}
Financial support by Russian Foundation for Basic Research (RFBR project No.~15-02-09366) is gratefully acknowledged.
\end{acknowledgments}

%
%%%% Appendix %%%%
%
\appendix
%
%%%% Material parameters %%%%
%
\section{Material parameters of MBBA}
\label{sec:appMBBA}
In the following we list the material parameters for MBBA which have been used for the theoretical analysis in this paper: 
dimensionless elastic constants $k_{11} =6.66$, $k_{22} =4.2$, $k_{33} =8.61$ in units of $k_0 =10^{-12}$~N;
dimensionless electric conductivities $\sigma_\parallel =1.5$, $\sigma_\perp =1$ in units of $\overline{\sigma_0}$;
dielectric constants $\epsilon_\parallel =4.72$, $\epsilon_\perp =5.25$;
dimensionless viscosity coefficients $\alpha_1 =-18.1$, $\alpha_2 =-110.4$, $\alpha_3 =-1.1$, $\alpha_4 =82.6$, $\alpha_5 =77.9$, $\alpha_6 =-33.6$ in units of $\alpha_0 = 10^{-3}$~Pa~s.
Here we have reproduced the data set introduced as MBBA~I in \cite{Bodenschatz:1988}, which has been used in a number of papers since (see, e.g., \cite{Kaiser:1993,Dressel:2002,Plaut:1999}).
In line with the experiments in Sec.~\ref{sec:experim} the cell thickness is fixed as $d = 25$~$\mu$m which yields $\tau_d = \alpha_0 d^2/(k_0 \pi^2) =0.0633$~s.
An excellent match of the theoretical and experimental values of $U_c(\omega)$, $q_c(\omega)$ in Fig.~\ref{CRITMBBA} is obtained by using $\overline{\sigma}_{0} = 1.25 \times 10^{-7}$~$(\Omega$~m)$^{-1}$ as fit parameter.
Thus $\tau_q =\epsilon_0 \epsilon_\perp/(\overline{\sigma}_0 \sigma_\perp)  = 0.372 \times 10^{-3}$~s, corresponding to a $Q = \tau_d/\tau_q \approx 170$.
%

%
%%%% Coefficients in the CAE %%%%
%
\section{Coefficients of  the coupled amplitude equations}
\label{sec:appCoeff}
In Table~\ref{tab:coeff} the coefficients of the CAE Eq.~(\ref{eq:ampAGPHI}), calculated for the material parameters given in Appendix~\ref{sec:appMBBA}, are listed for selected dimensionless frequencies $\omega^\prime =\omega\tau_q$; all coefficients show in general a monotonic dependence on $\omega^\prime$.
The calculations are based on the general formalism presented in \cite{Plaut:1999}; we use four $z$-modes, three Fourier modes in time [$\exp(\pm i n \omega t)$, $0 \le n \le 2$] and three Fourier modes in $x$ [$\exp(\pm i k q_c  x)$, $0 \le k \le 2$] to represent the fields $\phi$, $\bm n$, $\bm v$ in the NHE [see, e.g., Eqs.~(\ref{eq:nz}) and (\ref{eq:nzfour})].
We have tested that increasing the number of modes does not lead to visible modifications of the CAE stability diagram in Fig.~\ref{fig:STABAMP}.
\begin{longtable}{lrrrrr}
\caption{\label{tab:coeff}
Coefficients in the coupled amplitude equations (\ref{eq:ampAGPHI}).}
\\
\hline\hline
{Coeff.}& $\omega\tau_q=0.5$ & $\omega\tau_q=1.0$ & $\omega\tau_q=1.3$ & $\omega\tau_q=1.6$ & $\omega\tau_q=1.9$ \\ \hline\hline
$\tau_0$   & 1.9929 & 1.4317 & 1.0670 & 0.7167 & 0.3882 \\ 
$q_c$   & 1.5933 & 1.8081 & 2.0150 & 2.3190 & 2.8195 \\ \hline
$e_1$   & 1.5276 & 1.3964 & 1.2761 & 1.1213 & 0.9322 \\ 
$e_2$   & -0.0938 & -0.146 & -0.1758 & -0.1820 & -0.1251 \\ 
$e_3$   & 0.1549 & 0.0338 & -0.0435 & -0.1143 & -0.1505 \\ \hline
$r_1$   & 0.9127 & 0.7841 & 0.6828 & 0.5621 & 0.4165 \\ 
$r_2$   & 0.0870 & 0.1657 & 0.2136 & 0.2599 & 0.3066 \\ \hline
$a_1$   & -2.0205 & -1.9760 & -1.8665 & -1.6794 & -1.4258 \\ 
$a_2$   & -0.2818 & -0.2654 & -0.2266 & -0.1615 & -0.0793 \\ 
$a_3$   & -0.8905 & -0.9486 & -0.8949 & -0.7728 & -0.6392 \\ 
$a_4$   & -0.6358 & -0.5572 & -0.4818 & -0.3889 & -0.3035 \\ 
$a_5$   & -0.2519 & -0.3300 & -0.4072 & -0.5302 & -0.7769 \\ 
$a_6$   & -0.0103 & -0.0630 & -0.1005 & -0.1494 & -0.2398 \\ \hline
$s_1$   & 2.8581 & 2.3418 & 1.9601 & 1.5330 & 1.0195 \\ 
$s_2$   & 2.4493 & 2.0644 & 1.7670 & 1.4296 & 1.0591 \\ \hline
$b_1$   & 0.1219 & 0.2551 & 0.3632 & 0.5074 & 0.7312 \\ 
$b_2$   & 0.1847 & 0.4706 & 0.7023 & 1.0071 & 1.4693 \\ 
$b_3$   & 0.7318 & 0.4551 & 0.3112 & 0.2121 & 0.1586 \\ 
$b_4$   & -0.1815 & -0.2024 & -0.1930 & -0.1592 & -0.1014 \\ \hline
$\beta_1$  & 0.2303 & 0.4587 & 0.6959 & 1.0916 & 1.8842 \\ 
$\beta_2$  & -0.3560 & 0.0900 & 0.3841 & 0.6860 & 1.0498 \\
$\beta_3$  & 0.8861 & 1.1119 & 1.4289 & 2.0927 & 3.8605 \\ \hline
$\nu_a$   & 41.3000 & 41.3000 & 41.3000 & 41.3000 & 41.3000 \\ 
$\nu_b$   & 23.9500 & 23.9500 & 23.9500 & 23.9500 & 23.9500 \\ \hline
$q_1$   & 25.3168 & 28.6833 & 32.1739 & 37.5227 & 46.7913 \\ 
$q_2$   & -17.3895 & -17.4692 & -17.8390 & -18.6265 & -20.3171 \\ 
$q_3$   & -18.8906 & -19.2267 & -19.5125 & -19.7740 & -19.9609 \\ 
$q_4$   & 3.0407 & 3.5551 & 4.2481 & 5.4752 & 7.8299 \\ \hline
$\Gamma_G$  & 30.0423 & 38.0550 & 47.2205 & 63.2443 & 97.4061 \\ \hline
$\sigma_T$  & -0.0384 & -0.0384 & -0.0384 & -0.0384 & -0.0384 \\\hline 
$K_3$   & 0.0788 & 0.0788 & 0.0788 & 0.0788 & 0.0788 \\ 
$K_1$   & 0.0601 & 0.0603 & 0.0605 & 0.0607 & 0.0608 \\ \hline
$\Gamma_\varphi$ & 0.2973 & 0.3468 & 0.4088 & 0.5212 & 0.7600 \\ \hline
$g_\varphi$  & 0.0048 & 0.0048 & 0.0048 & 0.0048 & 0.0048 \\ \hline
$\gamma_1$  & 0.1520 & 0.1500 & 0.1535 & 0.1615 & 0.1759 \\ 
$\gamma_2$  & 0.1202 & 0.1268 & 0.1377 & 0.1583 & 0.1997 \\
\hline\hline
\end{longtable}
%

%\newpage
%%% References
%\bibliography{references}

\begin{thebibliography}{33}%
\makeatletter
\providecommand \@ifxundefined [1]{%
 \@ifx{#1\undefined}
}%
\providecommand \@ifnum [1]{%
 \ifnum #1\expandafter \@firstoftwo
 \else \expandafter \@secondoftwo
 \fi
}%
\providecommand \@ifx [1]{%
 \ifx #1\expandafter \@firstoftwo
 \else \expandafter \@secondoftwo
 \fi
}%
\providecommand \natexlab [1]{#1}%
\providecommand \enquote  [1]{``#1''}%
\providecommand \bibnamefont  [1]{#1}%
\providecommand \bibfnamefont [1]{#1}%
\providecommand \citenamefont [1]{#1}%
\providecommand \href@noop [0]{\@secondoftwo}%
\providecommand \href [0]{\begingroup \@sanitize@url \@href}%
\providecommand \@href[1]{\@@startlink{#1}\@@href}%
\providecommand \@@href[1]{\endgroup#1\@@endlink}%
\providecommand \@sanitize@url [0]{\catcode `\\12\catcode `\$12\catcode
  `\&12\catcode `\#12\catcode `\^12\catcode `\_12\catcode `\%12\relax}%
\providecommand \@@startlink[1]{}%
\providecommand \@@endlink[0]{}%
\providecommand \url  [0]{\begingroup\@sanitize@url \@url }%
\providecommand \@url [1]{\endgroup\@href {#1}{\urlprefix }}%
\providecommand \urlprefix  [0]{URL }%
\providecommand \Eprint [0]{\href }%
\providecommand \doibase [0]{http://dx.doi.org/}%
\providecommand \selectlanguage [0]{\@gobble}%
\providecommand \bibinfo  [0]{\@secondoftwo}%
\providecommand \bibfield  [0]{\@secondoftwo}%
\providecommand \translation [1]{[#1]}%
\providecommand \BibitemOpen [0]{}%
\providecommand \bibitemStop [0]{}%
\providecommand \bibitemNoStop [0]{.\EOS\space}%
\providecommand \EOS [0]{\spacefactor3000\relax}%
\providecommand \BibitemShut  [1]{\csname bibitem#1\endcsname}%
\let\auto@bib@innerbib\@empty
%</preamble>
\bibitem [{\citenamefont {Bodenschatz}\ \emph
  {et~al.}(1988{\natexlab{a}})\citenamefont {Bodenschatz}, \citenamefont
  {Zimmermann},\ and\ \citenamefont {Kramer}}]{Bodenschatz:1988}%
  \BibitemOpen
  \bibfield  {author} {\bibinfo {author} {\bibfnamefont {E.}~\bibnamefont
  {Bodenschatz}}, \bibinfo {author} {\bibfnamefont {W.}~\bibnamefont
  {Zimmermann}}, \ and\ \bibinfo {author} {\bibfnamefont {L.}~\bibnamefont
  {Kramer}},\ }\href@noop {} {\bibfield  {journal} {\bibinfo  {journal} {J.
  Phys. (France)}\ }\textbf {\bibinfo {volume} {49}},\ \bibinfo {pages} {1875}
  (\bibinfo {year} {1988}{\natexlab{a}})}\BibitemShut {NoStop}%
\bibitem [{\citenamefont {Buka}\ and\ \citenamefont
  {Kramer}(1996)}]{Buka:1996}%
  \BibitemOpen
  \bibinfo {editor} {\bibfnamefont {{\'A}.}~\bibnamefont {Buka}}\ and\ \bibinfo
  {editor} {\bibfnamefont {L.}~\bibnamefont {Kramer}},\ eds.,\ \href@noop {}
  {\emph {\bibinfo {title} {{Pattern Formation in Liquid Crystals}}}}\
  (\bibinfo  {publisher} {Springer},\ \bibinfo {address} {New York},\ \bibinfo
  {year} {1996})\BibitemShut {NoStop}%
\bibitem [{\citenamefont {Buka}\ \emph {et~al.}(2006)\citenamefont {Buka},
  \citenamefont {\'Eber}, \citenamefont {Pesch},\ and\ \citenamefont
  {Kramer}}]{Golovin:2006}%
  \BibitemOpen
  \bibfield  {author} {\bibinfo {author} {\bibfnamefont {{\'A}.}~\bibnamefont
  {Buka}}, \bibinfo {author} {\bibfnamefont {N.}~\bibnamefont {\'Eber}},
  \bibinfo {author} {\bibfnamefont {W.}~\bibnamefont {Pesch}}, \ and\ \bibinfo
  {author} {\bibfnamefont {L.}~\bibnamefont {Kramer}},\ }in\ \href@noop {}
  {\emph {\bibinfo {booktitle} {Self-Assembly, Pattern Formation and Growth
  Phenomena in Nano-Systems}}},\ \bibinfo {editor} {edited by\ \bibinfo
  {editor} {\bibfnamefont {A.~A.}\ \bibnamefont {Golovin}}\ and\ \bibinfo
  {editor} {\bibfnamefont {A.~A.}\ \bibnamefont {Nepomnyashchy}}}\ (\bibinfo
  {publisher} {Springer},\ \bibinfo {address} {Dordrecht},\ \bibinfo {year}
  {2006})\ pp.\ \bibinfo {pages} {55--82}\BibitemShut {NoStop}%
\bibitem [{\citenamefont {de~Gennes}(1974)}]{Gennes:1974}%
  \BibitemOpen
  \bibfield  {author} {\bibinfo {author} {\bibfnamefont {P.~G.}\ \bibnamefont
  {de~Gennes}},\ }\href@noop {} {\emph {\bibinfo {title} {The Physics of Liquid
  Crystals}}}\ (\bibinfo  {publisher} {Clarendon Press},\ \bibinfo {address}
  {Oxford},\ \bibinfo {year} {1974})\BibitemShut {NoStop}%
\bibitem [{\citenamefont {Chandrasekhar}(1977)}]{chandra:1977}%
  \BibitemOpen
  \bibfield  {author} {\bibinfo {author} {\bibfnamefont {S.}~\bibnamefont
  {Chandrasekhar}},\ }\href@noop {} {\emph {\bibinfo {title} {Liquid
  Crystals}}}\ (\bibinfo  {publisher} {University Press},\ \bibinfo {address}
  {Cambridge},\ \bibinfo {year} {1977})\BibitemShut {NoStop}%
\bibitem [{\citenamefont {Pleiner}\ and\ \citenamefont
  {Brand}(1996)}]{brand:1996}%
  \BibitemOpen
  \bibfield  {author} {\bibinfo {author} {\bibfnamefont {H.}~\bibnamefont
  {Pleiner}}\ and\ \bibinfo {author} {\bibfnamefont {H.~R.}\ \bibnamefont
  {Brand}},\ }in\ \href@noop {} {\emph {\bibinfo {booktitle} {Pattern Formation
  in Liquid Crystals}}},\ \bibinfo {editor} {edited by\ \bibinfo {editor}
  {\bibfnamefont {{\'A}.}~\bibnamefont {Buka}}\ and\ \bibinfo {editor}
  {\bibfnamefont {L.}~\bibnamefont {Kramer}}}\ (\bibinfo  {publisher}
  {Springer},\ \bibinfo {address} {New York},\ \bibinfo {year} {1996})\ pp.\
  \bibinfo {pages} {15--67}\BibitemShut {NoStop}%
\bibitem [{\citenamefont {Cross}\ and\ \citenamefont
  {Hohenberg}(1993)}]{Cross:1993}%
  \BibitemOpen
  \bibfield  {author} {\bibinfo {author} {\bibfnamefont {M.~C.}\ \bibnamefont
  {Cross}}\ and\ \bibinfo {author} {\bibfnamefont {P.~C.}\ \bibnamefont
  {Hohenberg}},\ }\href@noop {} {\bibfield  {journal} {\bibinfo  {journal}
  {Rev. Mod. Phys.}\ }\textbf {\bibinfo {volume} {65}},\ \bibinfo {pages} {851}
  (\bibinfo {year} {1993})}\BibitemShut {NoStop}%
\bibitem [{\citenamefont {Cross}\ and\ \citenamefont
  {Greenside}(2009)}]{CroGreen:2009}%
  \BibitemOpen
  \bibfield  {author} {\bibinfo {author} {\bibfnamefont {M.}~\bibnamefont
  {Cross}}\ and\ \bibinfo {author} {\bibfnamefont {H.}~\bibnamefont
  {Greenside}},\ }\href@noop {} {\emph {\bibinfo {title} {Pattern Formation and
  Dynamics in Nonequilibrium Systems}}}\ (\bibinfo  {publisher} {Cambridge
  University Press},\ \bibinfo {address} {New York},\ \bibinfo {year}
  {2009})\BibitemShut {NoStop}%
\bibitem [{\citenamefont {Bodenschatz}\ \emph
  {et~al.}(1988{\natexlab{b}})\citenamefont {Bodenschatz}, \citenamefont
  {Pesch},\ and\ \citenamefont {Kramer}}]{Bodenschatz2:1988}%
  \BibitemOpen
  \bibfield  {author} {\bibinfo {author} {\bibfnamefont {E.}~\bibnamefont
  {Bodenschatz}}, \bibinfo {author} {\bibfnamefont {W.}~\bibnamefont {Pesch}},
  \ and\ \bibinfo {author} {\bibfnamefont {L.}~\bibnamefont {Kramer}},\
  }\href@noop {} {\bibfield  {journal} {\bibinfo  {journal} {Physica D}\
  }\textbf {\bibinfo {volume} {32}},\ \bibinfo {pages} {135} (\bibinfo {year}
  {1988}{\natexlab{b}})}\BibitemShut {NoStop}%
\bibitem [{\citenamefont {Kramer}\ \emph {et~al.}(1990)\citenamefont {Kramer},
  \citenamefont {Bodenschatz},\ and\ \citenamefont {Pesch}}]{Bodencom:1990}%
  \BibitemOpen
  \bibfield  {author} {\bibinfo {author} {\bibfnamefont {L.}~\bibnamefont
  {Kramer}}, \bibinfo {author} {\bibfnamefont {E.}~\bibnamefont {Bodenschatz}},
  \ and\ \bibinfo {author} {\bibfnamefont {W.}~\bibnamefont {Pesch}},\
  }\href@noop {} {\bibfield  {journal} {\bibinfo  {journal} {Phys. Rev. Lett.}\
  }\textbf {\bibinfo {volume} {64}},\ \bibinfo {pages} {2588} (\bibinfo {year}
  {1990})}\BibitemShut {NoStop}%
\bibitem [{\citenamefont {Busse}\ and\ \citenamefont
  {Clever}(1979)}]{Busse:1979}%
  \BibitemOpen
  \bibfield  {author} {\bibinfo {author} {\bibfnamefont {F.~H.}\ \bibnamefont
  {Busse}}\ and\ \bibinfo {author} {\bibfnamefont {R.~M.}\ \bibnamefont
  {Clever}},\ }\href@noop {} {\bibfield  {journal} {\bibinfo  {journal} {J.
  Fluid Mech.}\ }\textbf {\bibinfo {volume} {91}},\ \bibinfo {pages} {319}
  (\bibinfo {year} {1979})}\BibitemShut {NoStop}%
\bibitem [{\citenamefont {Kaiser}\ and\ \citenamefont
  {Pesch}(1993)}]{Kaiser:1993}%
  \BibitemOpen
  \bibfield  {author} {\bibinfo {author} {\bibfnamefont {M.}~\bibnamefont
  {Kaiser}}\ and\ \bibinfo {author} {\bibfnamefont {W.}~\bibnamefont {Pesch}},\
  }\href@noop {} {\bibfield  {journal} {\bibinfo  {journal} {Phys. Rev. E}\
  }\textbf {\bibinfo {volume} {48}},\ \bibinfo {pages} {4510} (\bibinfo {year}
  {1993})}\BibitemShut {NoStop}%
\bibitem [{\citenamefont {Decker}\ and\ \citenamefont
  {Pesch}(1994)}]{Decker:1994}%
  \BibitemOpen
  \bibfield  {author} {\bibinfo {author} {\bibfnamefont {W.}~\bibnamefont
  {Decker}}\ and\ \bibinfo {author} {\bibfnamefont {W.}~\bibnamefont {Pesch}},\
  }\href@noop {} {\bibfield  {journal} {\bibinfo  {journal} {J. Phys. II
  (France)}\ }\textbf {\bibinfo {volume} {4}},\ \bibinfo {pages} {419}
  (\bibinfo {year} {1994})}\BibitemShut {NoStop}%
\bibitem [{\citenamefont {Plaut}\ \emph {et~al.}(1997)\citenamefont {Plaut},
  \citenamefont {Decker}, \citenamefont {Rossberg}, \citenamefont {Kramer},
  \citenamefont {Pesch}, \citenamefont {Belaidi},\ and\ \citenamefont
  {Ribotta}}]{Plaut:1997}%
  \BibitemOpen
  \bibfield  {author} {\bibinfo {author} {\bibfnamefont {E.}~\bibnamefont
  {Plaut}}, \bibinfo {author} {\bibfnamefont {W.}~\bibnamefont {Decker}},
  \bibinfo {author} {\bibfnamefont {A.~G.}\ \bibnamefont {Rossberg}}, \bibinfo
  {author} {\bibfnamefont {L.}~\bibnamefont {Kramer}}, \bibinfo {author}
  {\bibfnamefont {W.}~\bibnamefont {Pesch}}, \bibinfo {author} {\bibfnamefont
  {A.}~\bibnamefont {Belaidi}}, \ and\ \bibinfo {author} {\bibfnamefont
  {R.}~\bibnamefont {Ribotta}},\ }\href@noop {} {\bibfield  {journal} {\bibinfo
   {journal} {Phys. Rev. Lett.}\ }\textbf {\bibinfo {volume} {79}},\ \bibinfo
  {pages} {2367} (\bibinfo {year} {1997})}\BibitemShut {NoStop}%
\bibitem [{\citenamefont {Plaut}\ and\ \citenamefont
  {Pesch}(1999)}]{Plaut:1999}%
  \BibitemOpen
  \bibfield  {author} {\bibinfo {author} {\bibfnamefont {E.}~\bibnamefont
  {Plaut}}\ and\ \bibinfo {author} {\bibfnamefont {W.}~\bibnamefont {Pesch}},\
  }\href@noop {} {\bibfield  {journal} {\bibinfo  {journal} {Phys. Rev. E}\
  }\textbf {\bibinfo {volume} {59}},\ \bibinfo {pages} {1747} (\bibinfo {year}
  {1999})}\BibitemShut {NoStop}%
\bibitem [{\citenamefont {Dressel}\ \emph {et~al.}(2002)\citenamefont
  {Dressel}, \citenamefont {Joets}, \citenamefont {Pastur}, \citenamefont
  {Pesch}, \citenamefont {Plaut},\ and\ \citenamefont
  {Ribotta}}]{Dressel:2002}%
  \BibitemOpen
  \bibfield  {author} {\bibinfo {author} {\bibfnamefont {B.}~\bibnamefont
  {Dressel}}, \bibinfo {author} {\bibfnamefont {A.}~\bibnamefont {Joets}},
  \bibinfo {author} {\bibfnamefont {L.}~\bibnamefont {Pastur}}, \bibinfo
  {author} {\bibfnamefont {W.}~\bibnamefont {Pesch}}, \bibinfo {author}
  {\bibfnamefont {E.}~\bibnamefont {Plaut}}, \ and\ \bibinfo {author}
  {\bibfnamefont {R.}~\bibnamefont {Ribotta}},\ }\href@noop {} {\bibfield
  {journal} {\bibinfo  {journal} {Phys. Rev. Lett.}\ }\textbf {\bibinfo
  {volume} {88}},\ \bibinfo {pages} {024503} (\bibinfo {year}
  {2002})}\BibitemShut {NoStop}%
\bibitem [{\citenamefont {Krekhov}\ \emph {et~al.}(2011)\citenamefont
  {Krekhov}, \citenamefont {Pesch},\ and\ \citenamefont {Buka}}]{Krekhov:2011}%
  \BibitemOpen
  \bibfield  {author} {\bibinfo {author} {\bibfnamefont {A.}~\bibnamefont
  {Krekhov}}, \bibinfo {author} {\bibfnamefont {W.}~\bibnamefont {Pesch}}, \
  and\ \bibinfo {author} {\bibfnamefont {{\'A}.}~\bibnamefont {Buka}},\ }\href
  {\doibase 10.1103/PhysRevE.83.051706} {\bibfield  {journal} {\bibinfo
  {journal} {Phys. Rev. E}\ }\textbf {\bibinfo {volume} {83}},\ \bibinfo
  {pages} {051706} (\bibinfo {year} {2011})}\BibitemShut {NoStop}%
\bibitem [{\citenamefont {John}\ and\ \citenamefont
  {Stannarius}(2004)}]{John:2004}%
  \BibitemOpen
  \bibfield  {author} {\bibinfo {author} {\bibfnamefont {T.}~\bibnamefont
  {John}}\ and\ \bibinfo {author} {\bibfnamefont {R.}~\bibnamefont
  {Stannarius}},\ }\href@noop {} {\bibfield  {journal} {\bibinfo  {journal}
  {Phys. Rev. E.}\ }\textbf {\bibinfo {volume} {70}},\ \bibinfo {pages}
  {025202} (\bibinfo {year} {2004})}\BibitemShut {NoStop}%
\bibitem [{\citenamefont {Pietschmann}\ \emph {et~al.}(2010)\citenamefont
  {Pietschmann}, \citenamefont {John},\ and\ \citenamefont
  {Stannarius}}]{Pietschmann:2010}%
  \BibitemOpen
  \bibfield  {author} {\bibinfo {author} {\bibfnamefont {D.}~\bibnamefont
  {Pietschmann}}, \bibinfo {author} {\bibfnamefont {T.}~\bibnamefont {John}}, \
  and\ \bibinfo {author} {\bibfnamefont {R.}~\bibnamefont {Stannarius}},\
  }\href@noop {} {\bibfield  {journal} {\bibinfo  {journal} {Phys. Rev. E.}\
  }\textbf {\bibinfo {volume} {82}},\ \bibinfo {pages} {046215} (\bibinfo
  {year} {2010})}\BibitemShut {NoStop}%
\bibitem [{\citenamefont {Trainoff}\ and\ \citenamefont
  {Cannell}(2002)}]{Trainoff:2002}%
  \BibitemOpen
  \bibfield  {author} {\bibinfo {author} {\bibfnamefont {S.~P.}\ \bibnamefont
  {Trainoff}}\ and\ \bibinfo {author} {\bibfnamefont {D.~S.}\ \bibnamefont
  {Cannell}},\ }\href@noop {} {\bibfield  {journal} {\bibinfo  {journal} {Phys.
  Fluids}\ }\textbf {\bibinfo {volume} {14}},\ \bibinfo {pages} {1340}
  (\bibinfo {year} {2002})}\BibitemShut {NoStop}%
\bibitem [{\citenamefont {Pesch}\ and\ \citenamefont
  {Krekhov}(2013)}]{Pesch:2013}%
  \BibitemOpen
  \bibfield  {author} {\bibinfo {author} {\bibfnamefont {W.}~\bibnamefont
  {Pesch}}\ and\ \bibinfo {author} {\bibfnamefont {A.}~\bibnamefont
  {Krekhov}},\ }\href@noop {} {\bibfield  {journal} {\bibinfo  {journal} {Phys.
  Rev. E.}\ }\textbf {\bibinfo {volume} {87}},\ \bibinfo {pages} {052504}
  (\bibinfo {year} {2013})}\BibitemShut {NoStop}%
\bibitem [{\citenamefont {Rudroff}\ \emph {et~al.}(1999)\citenamefont
  {Rudroff}, \citenamefont {Frette},\ and\ \citenamefont
  {Rehberg}}]{Rudroff:1999}%
  \BibitemOpen
  \bibfield  {author} {\bibinfo {author} {\bibfnamefont {S.}~\bibnamefont
  {Rudroff}}, \bibinfo {author} {\bibfnamefont {V.}~\bibnamefont {Frette}}, \
  and\ \bibinfo {author} {\bibfnamefont {I.}~\bibnamefont {Rehberg}},\
  }\href@noop {} {\bibfield  {journal} {\bibinfo  {journal} {Phys. Rev. E}\
  }\textbf {\bibinfo {volume} {59}},\ \bibinfo {pages} {1814} (\bibinfo {year}
  {1999})}\BibitemShut {NoStop}%
\bibitem [{\citenamefont {Amm}\ \emph {et~al.}(1999)\citenamefont {Amm},
  \citenamefont {Stannarius},\ and\ \citenamefont {Rossberg}}]{Amm:1999}%
  \BibitemOpen
  \bibfield  {author} {\bibinfo {author} {\bibfnamefont {H.}~\bibnamefont
  {Amm}}, \bibinfo {author} {\bibfnamefont {R.}~\bibnamefont {Stannarius}}, \
  and\ \bibinfo {author} {\bibfnamefont {A.~G.}\ \bibnamefont {Rossberg}},\
  }\href@noop {} {\bibfield  {journal} {\bibinfo  {journal} {Physica D}\
  }\textbf {\bibinfo {volume} {126}},\ \bibinfo {pages} {171} (\bibinfo {year}
  {1999})}\BibitemShut {NoStop}%
\bibitem [{\citenamefont {Oikawa}\ \emph {et~al.}(2004)\citenamefont {Oikawa},
  \citenamefont {Hidaka},\ and\ \citenamefont {Kai}}]{Oikawa:2004}%
  \BibitemOpen
  \bibfield  {author} {\bibinfo {author} {\bibfnamefont {N.}~\bibnamefont
  {Oikawa}}, \bibinfo {author} {\bibfnamefont {Y.}~\bibnamefont {Hidaka}}, \
  and\ \bibinfo {author} {\bibfnamefont {S.}~\bibnamefont {Kai}},\ }\href@noop
  {} {\bibfield  {journal} {\bibinfo  {journal} {Phys. Rev. E}\ }\textbf
  {\bibinfo {volume} {70}},\ \bibinfo {pages} {066204} (\bibinfo {year}
  {2004})}\BibitemShut {NoStop}%
\bibitem [{\citenamefont {\'Eber}\ \emph {et~al.}(2012)\citenamefont {\'Eber},
  \citenamefont {Palomares}, \citenamefont {Salamon}, \citenamefont {Krekhov},\
  and\ \citenamefont {Buka}}]{Eber:2012}%
  \BibitemOpen
  \bibfield  {author} {\bibinfo {author} {\bibfnamefont {N.}~\bibnamefont
  {\'Eber}}, \bibinfo {author} {\bibfnamefont {L.~O.}\ \bibnamefont
  {Palomares}}, \bibinfo {author} {\bibfnamefont {P.}~\bibnamefont {Salamon}},
  \bibinfo {author} {\bibfnamefont {A.}~\bibnamefont {Krekhov}}, \ and\
  \bibinfo {author} {\bibfnamefont {{\'A}.}~\bibnamefont {Buka}},\ }\href@noop
  {} {\bibfield  {journal} {\bibinfo  {journal} {Phys. Rev. E}\ }\textbf
  {\bibinfo {volume} {86}},\ \bibinfo {pages} {021702} (\bibinfo {year}
  {2012})}\BibitemShut {NoStop}%
\bibitem [{\citenamefont {Ribotta}\ \emph {et~al.}(1986)\citenamefont
  {Ribotta}, \citenamefont {Joets},\ and\ \citenamefont {Lei}}]{Ribotta:1986}%
  \BibitemOpen
  \bibfield  {author} {\bibinfo {author} {\bibfnamefont {R.}~\bibnamefont
  {Ribotta}}, \bibinfo {author} {\bibfnamefont {A.}~\bibnamefont {Joets}}, \
  and\ \bibinfo {author} {\bibfnamefont {L.}~\bibnamefont {Lei}},\ }\href@noop
  {} {\bibfield  {journal} {\bibinfo  {journal} {Phys. Rev. Lett.}\ }\textbf
  {\bibinfo {volume} {56}},\ \bibinfo {pages} {1595} (\bibinfo {year}
  {1986})}\BibitemShut {NoStop}%
\bibitem [{\citenamefont {Joets}\ and\ \citenamefont
  {Ribotta}(1991)}]{Joets:1991}%
  \BibitemOpen
  \bibfield  {author} {\bibinfo {author} {\bibfnamefont {A.}~\bibnamefont
  {Joets}}\ and\ \bibinfo {author} {\bibfnamefont {R.}~\bibnamefont
  {Ribotta}},\ }\href@noop {} {\bibfield  {journal} {\bibinfo  {journal} {J.
  Stat. Phys.}\ }\textbf {\bibinfo {volume} {64}},\ \bibinfo {pages} {981}
  (\bibinfo {year} {1991})}\BibitemShut {NoStop}%
\bibitem [{\citenamefont {Nasuno}\ \emph {et~al.}(1992)\citenamefont {Nasuno},
  \citenamefont {Sasaki}, \citenamefont {Kai},\ and\ \citenamefont
  {Zimmermann}}]{Nasuno:1992}%
  \BibitemOpen
  \bibfield  {author} {\bibinfo {author} {\bibfnamefont {S.}~\bibnamefont
  {Nasuno}}, \bibinfo {author} {\bibfnamefont {O.}~\bibnamefont {Sasaki}},
  \bibinfo {author} {\bibfnamefont {S.}~\bibnamefont {Kai}}, \ and\ \bibinfo
  {author} {\bibfnamefont {W.}~\bibnamefont {Zimmermann}},\ }\href@noop {}
  {\bibfield  {journal} {\bibinfo  {journal} {Phys. Rev. A}\ }\textbf {\bibinfo
  {volume} {46}},\ \bibinfo {pages} {4954} (\bibinfo {year}
  {1992})}\BibitemShut {NoStop}%
\bibitem [{\citenamefont {Rudroff}\ \emph {et~al.}(1998)\citenamefont
  {Rudroff}, \citenamefont {Zhao}, \citenamefont {Kramer},\ and\ \citenamefont
  {Rehberg}}]{Rudroff:1998}%
  \BibitemOpen
  \bibfield  {author} {\bibinfo {author} {\bibfnamefont {S.}~\bibnamefont
  {Rudroff}}, \bibinfo {author} {\bibfnamefont {H.}~\bibnamefont {Zhao}},
  \bibinfo {author} {\bibfnamefont {L.}~\bibnamefont {Kramer}}, \ and\ \bibinfo
  {author} {\bibfnamefont {I.}~\bibnamefont {Rehberg}},\ }\href@noop {}
  {\bibfield  {journal} {\bibinfo  {journal} {Phys. Rev. Lett.}\ }\textbf
  {\bibinfo {volume} {81}},\ \bibinfo {pages} {4144} (\bibinfo {year}
  {1998})}\BibitemShut {NoStop}%
\bibitem [{\citenamefont {Rossberg}\ and\ \citenamefont
  {Kramer}(1998)}]{Rossberg2:1998}%
  \BibitemOpen
  \bibfield  {author} {\bibinfo {author} {\bibfnamefont {A.~G.}\ \bibnamefont
  {Rossberg}}\ and\ \bibinfo {author} {\bibfnamefont {L.}~\bibnamefont
  {Kramer}},\ }\href@noop {} {\bibfield  {journal} {\bibinfo  {journal}
  {Physica D}\ }\textbf {\bibinfo {volume} {115}},\ \bibinfo {pages} {19}
  (\bibinfo {year} {1998})}\BibitemShut {NoStop}%
\bibitem [{\citenamefont {Rossberg}(1998)}]{Rossberg:1998}%
  \BibitemOpen
  \bibfield  {author} {\bibinfo {author} {\bibfnamefont {A.~G.}\ \bibnamefont
  {Rossberg}},\ }\href@noop {} {Ph.D. thesis},\ \bibinfo  {school}
  {Universit{\"a}t Bayreuth} (\bibinfo {year} {1998})\BibitemShut {NoStop}%
\bibitem [{\citenamefont {Komineas}\ \emph {et~al.}(2003)\citenamefont
  {Komineas}, \citenamefont {Zhao},\ and\ \citenamefont
  {Kramer}}]{Komineas:2003}%
  \BibitemOpen
  \bibfield  {author} {\bibinfo {author} {\bibfnamefont {S.}~\bibnamefont
  {Komineas}}, \bibinfo {author} {\bibfnamefont {H.}~\bibnamefont {Zhao}}, \
  and\ \bibinfo {author} {\bibfnamefont {L.}~\bibnamefont {Kramer}},\
  }\href@noop {} {\bibfield  {journal} {\bibinfo  {journal} {Phys. Rev. E}\
  }\textbf {\bibinfo {volume} {67}},\ \bibinfo {pages} {031701} (\bibinfo
  {year} {2003})}\BibitemShut {NoStop}%
\bibitem [{\citenamefont {Batyrshin}\ \emph {et~al.}(2012)\citenamefont
  {Batyrshin}, \citenamefont {Krekhov}, \citenamefont {Scaldin},\ and\
  \citenamefont {Delev}}]{Batyrshin:2012}%
  \BibitemOpen
  \bibfield  {author} {\bibinfo {author} {\bibfnamefont {E.~S.}\ \bibnamefont
  {Batyrshin}}, \bibinfo {author} {\bibfnamefont {A.~P.}\ \bibnamefont
  {Krekhov}}, \bibinfo {author} {\bibfnamefont {O.~A.}\ \bibnamefont
  {Scaldin}}, \ and\ \bibinfo {author} {\bibfnamefont {V.~A.}\ \bibnamefont
  {Delev}},\ }\href@noop {} {\bibfield  {journal} {\bibinfo  {journal} {JETP}\
  }\textbf {\bibinfo {volume} {114}},\ \bibinfo {pages} {1052} (\bibinfo {year}
  {2012})}\BibitemShut {NoStop}%
\end{thebibliography}


%
\end{document}